\documentclass[reprint,aps,pra,superscriptaddress,showpacs,nofootinbib]{revtex4-2}
\usepackage{blindtext}
\usepackage{graphicx}
\usepackage{dcolumn}
\usepackage{bm}
\usepackage{color}
\usepackage{ulem}
\usepackage{soul}
\usepackage{siunitx}
\usepackage{amsmath}
\usepackage[english]{babel}
\usepackage[symbol]{footmisc}
\usepackage{mathrsfs}
\usepackage{comment}
\usepackage{appendix}

\begin{document}
\title{Toward quantum-noise-limited interferometric measurements of optical nonlinearity in vacuum}

\author{A. Aras}
\email{ali.aras@ijclab.in2p3.fr}
\author{A. E. Kraych}
\author{X. Sarazin}
\author{E. Baynard}
\author{F. Couchot}
\author{M. Pittman}
\affiliation{Université Paris-Saclay, CNRS/IN2P3, IJCLab, 91405 Orsay, France}

\begin{abstract}
\noindent \textbf{Abstract.} 
Quantum Electrodynamics predicts that the vacuum must behave as a nonlinear optical medium: the vacuum optical index should increase when it is stressed by intense electromagnetic fields. The DeLLight (Deflection of Light by Light) project aims to measure it by using intense and ultra-short laser pulses.
The experiment uses a Sagnac interferometer to amplify the tiny deflection signal of a low-intensity probe pulse crossing the vacuum refractive-index gradient produced by an external high-intensity pump pulse.
The measurement of the amplified signal by a CCD camera requires a high spatial resolution, which is limited by the ultimate quantum noise of the CCD. 
However, interferometric phase noise induced by the mechanical vibrations of the interferometer is also amplified and degrades spatial resolution.
To overcome this, we propose a new method named High-Frequency Phase Noise Suppression (HFPNS), based on the addition of a delayed replica (5 ns) of the probe pulse. The delayed pulse, which is not affected by the pump but is subject to the same vibration noise, enables offline subtraction of correlated phase noise.
In this work, we present an experimental proof-of-concept on a prototype interferometer operating with a limited amplification factor ($\mathcal{A}\simeq25$), about 10 times smaller than the required value of the final experiment.
We have succeeded in reducing phase noise by a factor of 40, resulting in a residual noise level 2.3 times higher than the expected quantum noise. The residual noise is linked to delay-line instabilities and incident beam pointing fluctuations present during these tests. 
This result validates HFPNS as a robust method for future quantum-noise-limited interferometric measurements of vacuum optical nonlinearity, though additional stabilization and higher interferometric amplification are still needed.

\end{abstract}
\maketitle

\section{Introduction}

Classical electrodynamics treats the vacuum as a simple, passive, and empty space. Therefore, the Maxwell's equations are written linear in a vacuum preventing any interactions between electromagnetic fields. It is characterized by two universal constants: the electric permittivity $\epsilon_0$ and the magnetic permeability $\mu_0$. The speed of light, $c$, which is directly related to those quantities, is also a universal constant. 

In contrast, Quantum Electrodynamics (QED) proposes the concept of a dynamic vacuum, where virtual particle-antiparticle pairs are constantly created and annihilated due to quantum fluctuations. The presence of virtual charged particles makes {\it light-by-light} interaction possible. Therefore, in the presence of external electromagnetic fields, the vacuum should behave as a nonlinear optical medium, leading to a decrease in the speed of light in vacuum. This phenomenon was first predicted by Euler and Heisenberg \cite{euler1935streuung, heisenberg1936folgerungen}, and later formulated within the QED framework by Schwinger \cite{schwinger1951gauge} as photon-photon scattering.

From a corpuscular perspective, photon-photon scattering has already been observed in several experiments, firstly at the Stanford Linear Accelerator Center (SLAC) \cite{burke1997positron} with real photons and later at the Large Hadron Collider (LHC) \cite{atlas2017evidence, sirunyan2019evidence} with quasi-real photons. In these cases, the vacuum appears only in the virtual electron-positron pair exchange, but its fundamental electromagnetic properties remain unchanged, i.e. no modification of the vacuum electromagnetic constants $\epsilon_0$, $\mu_0$, and c. 

However, the optical nonlinearity in vacuum, a pure classical undulatory process with a modification of the light velocity at macroscopic scale, has never been observed due to the need for ultra-high electric or magnetic fields, as well as ultra-sensitive measurement techniques. The experimental efforts on this scope focus mainly on testing vacuum magnetic birefringence in the presence of an external magnetic field \cite{q&aexperiment, ovalexperiment}. The best sensitivity reached so far has been achieved by the PVLAS experiment, measuring the vacuum magnetic birefringence 7 times higher than the expected QED signal (in 1$\sigma$ confidence level), after 100 days of collected data \cite{EJLLI20201}. New birefringence projects have been proposed in the recent years to CERN and XFEL to improve the sensitivity \cite{refId0, XU2020164553}.

The DeLLight experiment is an ongoing experiment designed to measure for the first time the optical non-linearity of the vacuum by using ultra-intense femtosecond laser pulses delivered by the LASERIX facility at IJCLab (Paris-Saclay University). The experimental method involves measuring the refraction of a low-intensity probe laser pulse after crossing a vacuum optical index gradient imprinted by its nonlinear interaction with an external intense pump laser pulse. Since the QED-predicted deflection is extremely small, a Sagnac interferometer is employed to amplify the deflection signal. In the initial phase, the proof of concept of the DeLLight experiment was achieved by measuring the deflection of light by light in air, with low-intensity external pump pulses, demonstrating the amplification of the deflection signal through interferometric measurement \cite{PhysRevA.109.053510, mailliet2024performance}.

In addition to the signal amplification, the detection of the QED deflection signal in vacuum requires measuring the barycenter position of the interference signal with high spatial accuracy. However, the spatial resolution is highly limited by the phase noise induced by mechanical vibrations of the interferometer. In order to suppress the vibrational phase noise, we have developed a novel method, named High-Frequency Phase Noise Suppression (HFPNS), which consists of splitting the incident probe laser pulse before entering the interferometer, into two identical pulses, one being delayed by few nanoseconds. The delayed pulse is then used to monitor and suppress the off-line interferometric phase noise considering the interferometric noise being the same for both the prompt and delayed pulses. 

In this article, we first present the DeLLight experiment and the interferometric amplification method, detailing the different origins of noise limiting the spatial resolution. We then explain the HFPNS method, and describe the experimental setup developed to test and validate the method. Finally, the results obtained on a prototype interferometric configuration without a pump beam are presented, demonstrating a significant suppression of the correlated vibrational phase noise together with the associated off-line data analysis procedure. The present work corresponds to a methodological validation of the HFPNS correction principle in a prototype interferometric configuration.


\section{DeLLight Experiment}

\subsection{Interferometric measurement}

\begin{figure}[h]
\centering
\includegraphics[width=1\linewidth]{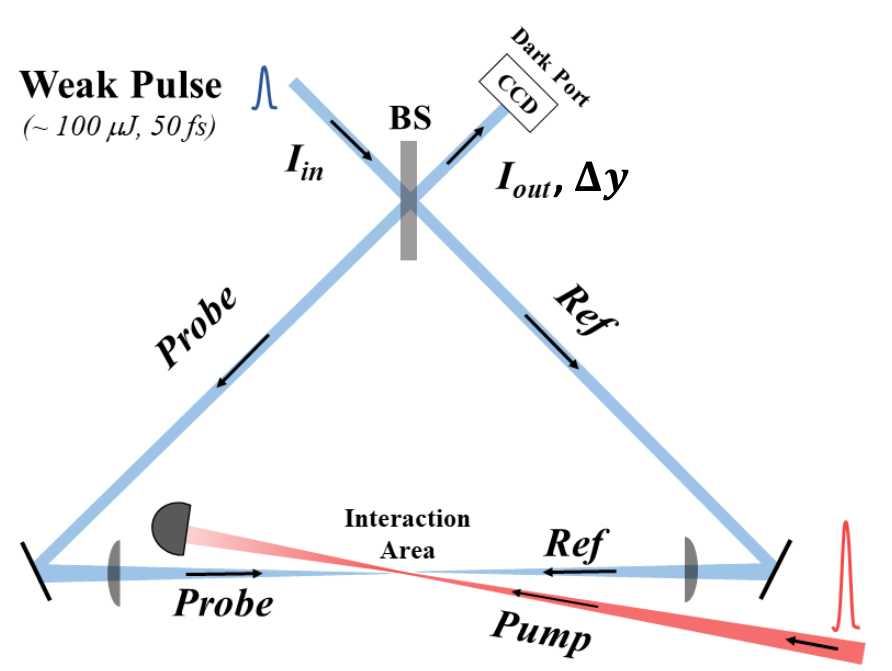}
\caption{Schematic layout of the Sagnac interferometer used in the DeLLight experiment. A weak probe pulse is injected into the interferometer and split by the beamsplitter (BS) into two counter-propagating pulses, referred to as the probe and reference beams. The external intense pump pulse intersects only the probe beam in the interaction region, inducing a small QED-predicted transverse deflection. At the dark output of the interferometer, this deflection is converted into an amplified displacement $\Delta y$ of the interference intensity profile, which is recorded by the CCD camera.}
\label{fig:sagnac-scheme-pump}
\end{figure}

The DeLLight interferometric method to measure the deflection of a probe laser pulse by an intense pump laser pulse is as follows, and schematized in Figure~\ref{fig:sagnac-scheme-pump}. A low-intensity laser pulse, of intensity $I_{in}$, is sent into a Sagnac interferometer via a 50/50 beam splitter, {\it BS}, generating two daughter pulses (the probe and reference pulses) that circulate in opposite directions around the interferometer. Both pulses are focused via two optical lenses in the interaction area with a minimum waist at focus $w_0$. 

The two counter-propagating pulses are in opposite phase in the dark output of the interferometer and therefore interfere destructively. Due to the small asymmetry between the reflection and transmission coefficients of the beam splitter, the destructive interference is never perfect, and a residual interference signal, of intensity $I_{out}$, is captured by a CCD camera in the dark output. The degree of destructive interference, i.e. the extinction of the residual interference signal, is characterized by the extinction factor $\mathcal{F}$, defined as $\mathcal{F} = I_{out}/I_{in}$ where $I_{in}$ is the incident probe intensity entering the interferometer. A smaller extinction factor corresponds to a deeper destructive interference and therefore to a larger interferometric amplification of the deflection signal.

The intense pump pulse of energy $E_{pump}$ is now focused into the same interaction area with a minimum waist at focus $W_0$, in time coincidence and counter-propagating configuration with respect to the probe pulse. Its nonlinear coupling with the probe in vacuum generates an optical index gradient proportional to the intensity profile of the pump. The probe is then refracted due to this emergent index gradient by a deflection angle $\delta \theta_{\mathrm{QED}}$, while the reference pulse is not in time coincidence with the pump and is therefore unaffected. After recollimation by the second lens, the refracted probe pulse is vertically shifted with respect to the unrefracted reference pulse by an average distance of $\delta y_{\mathrm{QED}} = f \times \delta \theta_{\mathrm{QED}} $, where $f$ is the focal length. 

As the deflected probe pulse interferes destructively with the unperturbed reference pulse, the transverse intensity profile of the interference signal in the dark output is then vertically shifted by a distance $\Delta y_{\mathrm{QED}}$, which is amplified as compared to the direct would-be signal $\delta y_{\mathrm{QED}}$ obtained when using a standard pointing method~\cite{PhysRevA.109.053510}. 

The amplification factor, defined as $\mathcal{A} = \Delta y_{\mathrm{QED}}/\delta y_{\mathrm{QED}}$, scales as the inverse square root of the extinction factor, i.e. $\mathcal{F}^{-1/2}$. The greater the extinction is, the greater the amplified signal is. As studied in~\cite{mailliet2024performance}, the optimum extinction factor $\mathcal{F}= 4 \times 10^{-6}$ has been experimentally demonstrated, corresponding to an amplification $\mathcal{A} = 250$. With the available energy of 2.5~J per pulses, delivered by the LASERIX facility, focused to $W_0=5$~$\mu$m, resulting in intensity on the order of $2 \times 10^{20}$ W/cm$^2$, the expected QED signal is  $\Delta y_{\mathrm{QED}}\approx15$~pm.

\subsection{Spatial resolution}

The deflection signal is measured by alternating laser shots with and without interaction between the pump and the probe, referred to as “ON” and “OFF” measurements, respectively. For each laser shot, the resulting interference pattern is recorded by a CCD camera at the dark output of the interferometer. We extract the vertical barycenters of the interference intensity profile for successive ON and OFF measurements, which we name $\bar{y}^{\mathrm{ON}}(i)$ and $\bar{y}^{\mathrm{OFF}}(i)$, respectively with $i$ representing the measurement number. The deflection signal $\Delta y (i)$ of the $i^{\mathrm{th}}$ "ON-OFF" measurement, corresponding to a shift of the barycenter due to the interaction with the pump, is then defined as
\begin{eqnarray}
\Delta y (i) = \bar{y}^{\mathrm{ON}}(i) - \bar{y}^{\mathrm{OFF}}(i).
\end{eqnarray}

The measured $\Delta y (i)$ values have a certain distribution characterized by its mean value $\langle{\Delta y}\rangle$ and its standard deviation $\sigma_{y}$, which we hereafter refer to as the {\it spatial resolution}. Collecting $N$ ON-OFF measurements, the statistical error (one standard deviation) of the measured average value $\langle{\Delta y}\rangle$ is then equal to $\sigma_y/\sqrt{N}$. So, regardless the available intensity of the pump pulses at focus and the degree of extinction of the interferometer, the capacity of detecting the QED signal requires a large number of measurements $N$, but also, and above all, an excellent spatial resolution $\sigma_y$.

The spatial resolution $\sigma_{y}$ is inherently limited by the intrinsic shot noise (quantum noise), related to the statistical fluctuations of the average number of photo-electrons detected by the CCD camera. As detailed in \cite{PhysRevA.103.023524}, the shot noise is independent of the beam width and it scales as $d_{pix}/\sqrt{N_{c}}$, where $d_{pix}$ is the side length of each square pixels and $N_{c}$ is the maximum number of detected photo-electrons per pixel before saturation, referred to as the full well capacity of the CCD camera. As measured in~\cite{mailliet2024performance}, a spatial resolution of $\sigma_y = 15$~nm can be obtained with the most appropriate commercial CCD camera. Assuming that the spatial resolution is only limited by the shot noise ($\sigma_y = 15$~nm), and with the 10~Hz repetition rate of the LASERIX facility, the expected sensitivity at 1$\sigma$ confidence level to measure the QED signal could be reached after about 4~days of collected data.

However, significant beam pointing fluctuations are present, leading to fluctuations $\delta y_{BP}$ of the barycenter of the intensity profile of the interference signal. Depending on the environmental conditions, these fluctuations are typically on the order of a few micrometers which is three orders of magnitude larger than the shot noise. Such fluctuations induce at first order a global shift of the barycenter of the interference intensity profile and limits the extraction of the QED-induced deflection signal. 

As demonstrated in~\cite{mailliet2024performance}, when operating the interferometer without amplification, any beam pointing fluctuation of the interference signal can be monitored in real-time. These fluctuations are measured through their correlation with the back reflection spot originating from the rear surface of the beam splitter, which provide a direct image of the incident beam on the same CCD camera. Since the back reflection experiences the same beam pointing perturbation as the interference signal, it can be used as an auxiliary reference to suppress these fluctuations during the off-line analysis procedure.

In contrast, when the interferometer operates in the amplification regime, the spatial resolution becomes significantly degraded by the interferometric phase noise induced by mechanical vibrations of the interferometer. These vibrations produce a relative lateral displacement $\delta y_{\Phi}$ between the two daughter pulses of the interferometer, namely the probe and reference beams, at the dark output. This relative displacement leads to an amplified displacement $\Delta y_{\Phi}$ of the interference intensity profile and therefore to an amplified noise contribution.

In this regime, the barycenter of the interference intensity profile is no longer directly correlated with the barycenter of the back reflection spot. In addition, this phase noise becomes even more dominant when increasing the analysis window size over which the barycenter is calculated, although large analysis windows are required to achieve high signal-detection efficiency, as discussed in Section~\ref{results}.

As a consequence, the current level of interferometric phase noise constitutes one of the main limitations for reaching the ultimate shot-noise-limited spatial resolution. It has been shown in~\cite{mailliet2024performance} that the present vibration level would need to be reduced by approximately two orders of magnitude in order to achieve the ultimate shot-noise limit. Reaching such a level of mechanical stability would require the development of a technically challenging and costly ultra-stable isolation system.

Here, an alternative method has been developed, named as High-Frequency Phase Noise Suppression (HFPNS), with the aim of measuring directly the displacement induced by the phase noise and after suppressing it off-line. It simply consists of splitting the incident probe laser pulse before entering the interferometer, into two identical pulses, one being delayed by few nanoseconds. Due to the small delay time, the vibration noise being the same for both the prompt and delayed pulses. The delayed pulse is then used to monitor and suppress off-line the interferometric phase noise. Details of the HFPNS method, description of the setup developed to demonstrate and validate the method in a prototype and results are given in the next two sections, section \ref{HFPNS}, and \ref{results}.
 
\section{High Frequency Phase Noise Suppression (HFPNS) Method\label{HFPNS}}

\subsection{Description of the HFPNS Method}

\begin{figure}[h]
    \centering
    \includegraphics[width=1\linewidth]{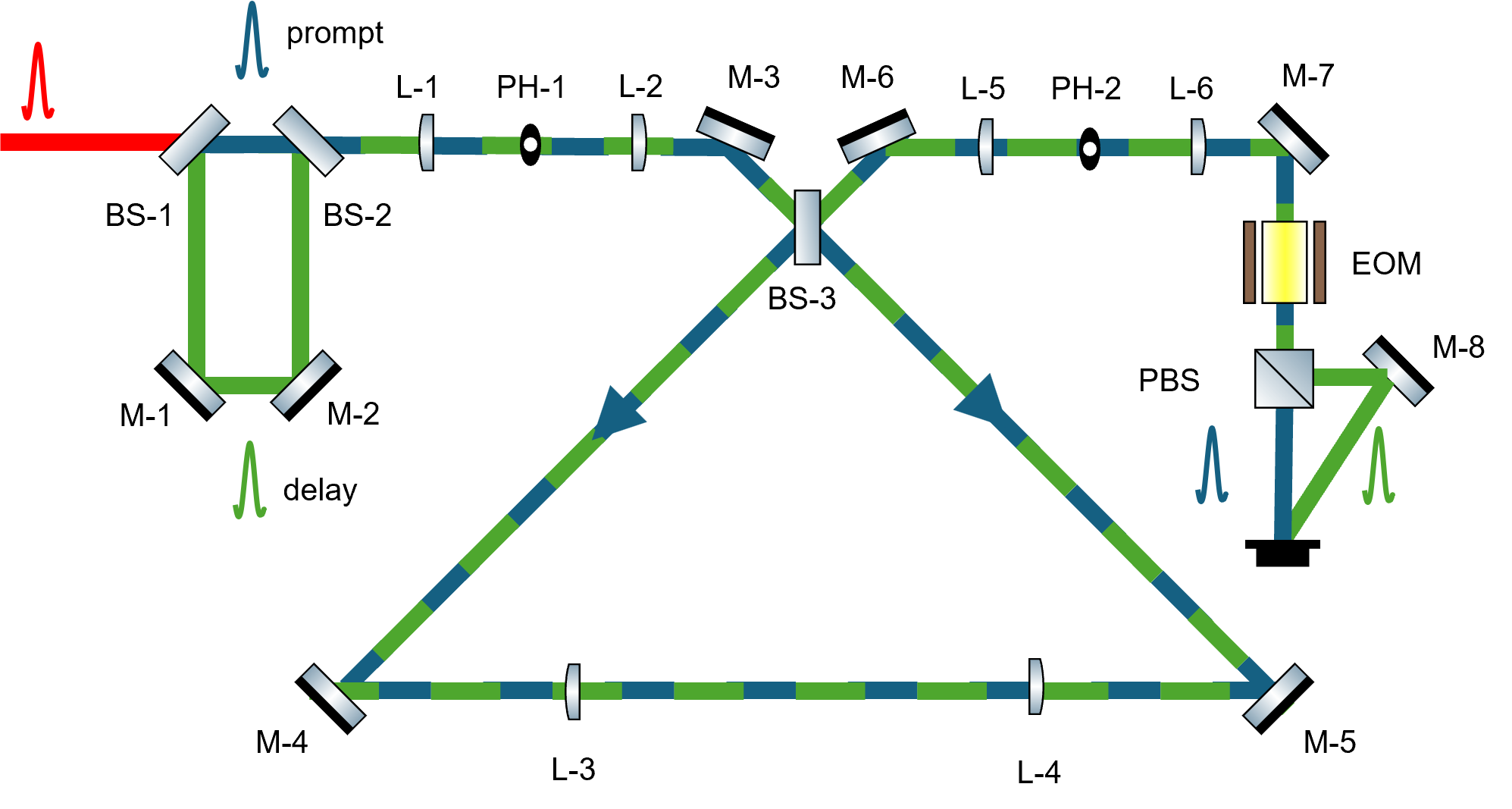}
    \caption{A simplified scheme of the optical design of DeLLight experiment with HFPNS method setup.}
    \label{fig:HFPNS-schema}
\end{figure}

The optical setup of the HFPNS method is schematically shown in Figure~\ref{fig:HFPNS-schema}. It consists of splitting the incident probe beam into two identical pulses. One pulse is transmitted directly, while the other one being delayed by about $5~ns$ using a delay line composed of two beam splitters (BS-1 and BS-2) and two mirrors (M1 and M2). 

The two time-separated beams are then directed to a spatial filter composed of a telescope of two lenses (L-1 and L-2) of same focal length 200~mm, and a pinhole (PH-1) with a diameter of 300~$\mu$m, ensuring that they both have similar beam profile close to a Gaussian profile, before entering inside the Sagnac interferometer.

In the current HFPNS interferometer prototype, a 50/50 femtosecond p-pol beam splitter BS-3 (Semrock FS01-BSTiS-5050P-25.5) is used to generate two daughter pulses, as explained in the previous section. The two dielectric mirrors (M-4 and M-5) are also positioned at $22.5^{\circ}$ incident angle, forming a right-angle triangular interferometer. A telescope of two best form spherical lenses (L-3 and L-4) of focal length 100 mm (Thorlabs LBF254-100-B) is placed between these two mirrors in order to focus both pulses into the interaction area. The incident beam waist is $w \simeq 1$~mm, corresponding to a minimum waist $w_{0} \simeq 25 \mu m$ at focus.

At the interferometer output, destructive interference pattern is separately obtained for both the prompt and delay pulses. Any scattered light from the interferometer is filtered by using a spatial filter consisting of a telescope with two 200~mm focal length lenses (L-5 and L-6) followed by a pinhole (PH-2) of 300~$\mu$m diameter, ensuring a clean Gaussian intensity profile in the interferometer output.

Finally, an electro-optic modulator (EOM) (KD*P Pockels Cell, Leysop Ltd) at 7.2kV is utilized in order to change the polarization of only the delay beam in high frequency (200 MHz) from pure p-state to s-state. The polarizing beam splitter cube (PBS) is then used to separate spatially these two prompt and delay pulses from the same beam line. Finally, the output images are collected on the same CCD camera (Basler acA3088-16gm, pixel size $5.84 \times 5.84~\mu$m$^2$) via a mirror (M-8). The efficiency of this separation on the CCD is measured by the intensity ratio of the delay beam on the prompt spot which is measured as 1:20.

Considering the small delay time of $5~ns$, the phase noise induced by the vibrations inside the interferometer, and corresponding to the lateral shift $\Delta y_{\Phi}$ of the interference intensity profiles, is then found identical for both the prompt $I_{out}^P(y)$ and the delay $I_{out}^D(y)$, while they are both affected by the same beam pointing fluctuations (neglecting the relative vibrations of the delay line), and are therefore correlated.

In contrast, thanks to their time coincidence, only the prompt pulse experiences the pump-induced deflection, which generates a shift $\Delta y_{\mathrm{QED}}$ in its interference intensity profile. The resulting intensity distributions for the prompt and delayed pulses can therefore be written as
\begin{eqnarray}
I_{out}^P(y) & = & \mathcal{F} \times I_{in}(y+\Delta y_{BP}+\Delta y_{\Phi}+\Delta y_{\mathrm{QED}}) \label{eqn:prompt-out-signal} \\
I_{out}^D(y) & = & \mathcal{F} \times I_{in}(y+\Delta y_{BP}+\Delta y_{\Phi}) \label{eqn:delay-out-signal} 
\end{eqnarray}

This configuration allows the delayed pulse to be used for the simultaneous measurement of beam pointing fluctuations and interferometric phase noise. Their contributions in the estimated barycenter, then, can be subtracted off-line in the analysis procedure, enabling the extraction of the pure QED-induced deflection signal $\Delta y_{\mathrm{QED}}$.

\subsection{Extinction factor}

\begin{figure}[h]
  \centering
  \includegraphics[width=1\linewidth]{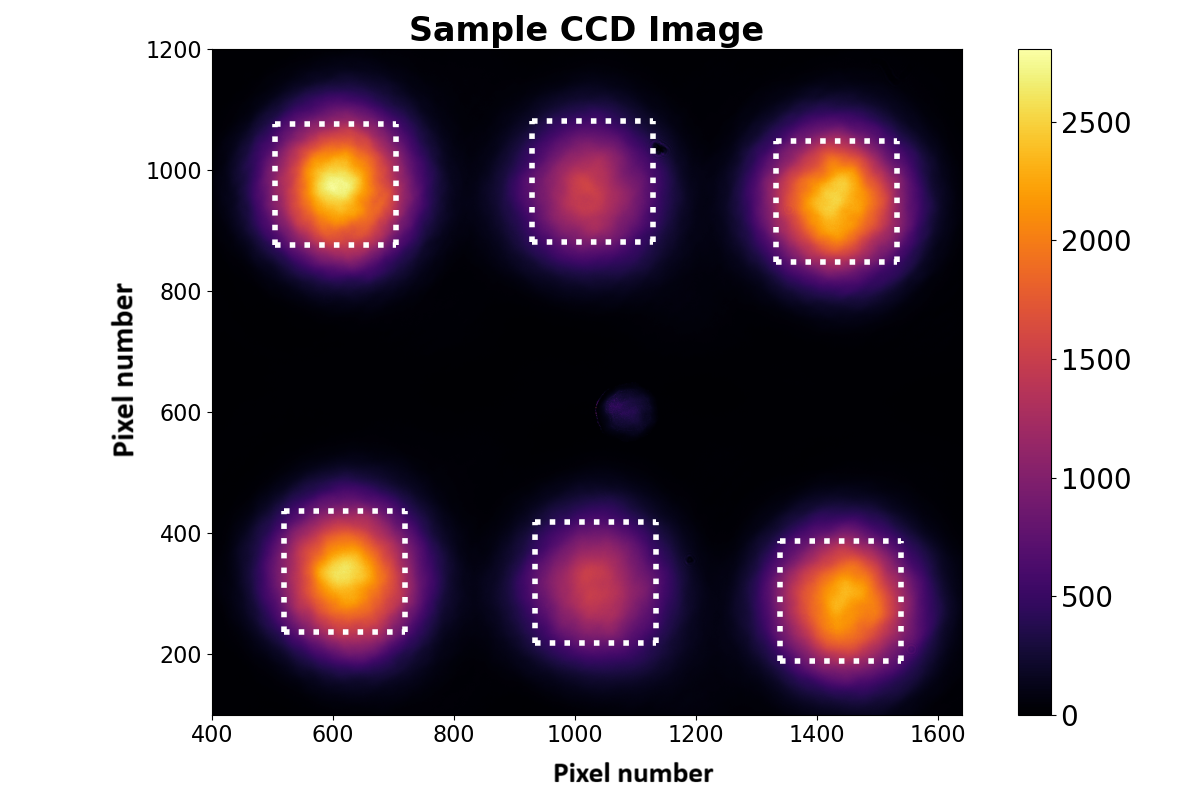}
  \caption{A CCD image of the intensity profiles of the prompt (top) and delay (bottom) probe pulses, recorded in the dark output of the interferometer. The interference signals are located in the central part. The two opposite lateral spots correspond to the back reflections on the rear side of the beam splitter. The white dotted square areas correspond to the Region of Interest (RoI) used to calculate the barycenters. 
  }
  \label{fig:sample-image-hfpns}
\end{figure}

A typical transverse intensity profile recorded by the CCD camera at the dark output of the HFPNS interferometer is shown in Figure~\ref{fig:sample-image-hfpns}. The upper row of three spots corresponds to the prompt pulse, while the lower row represents the delayed pulse. Within each row, the central spot corresponds to the interference signal. The two additional spots observed on opposite lateral sides originate from back reflections on the rear surface of the beam splitter. Their respective intensities are $I_{AR,1} = I_{AR,2} = (R_{AR}/2) I_{in}$, where $R_{AR} \simeq 10^{-3}$ is the back reflection coefficient of the beam splitter.

In order to reach the ultimate spatial resolution, high photon statistics are required in the interference signal intensity profile. This can be achieved by increasing the intensity of the incident probe pulse. However, with the currently available beam splitter, the achievable extinction factor becomes mainly limited by the back reflections, which can eventually saturate the CCD camera. To overcome this limitation, the extinction is adjusted such that the interference-signal intensity remains of the same order of magnitude as the back reflection intensities. This tuning is obtained by slightly rotating the interferometer beam splitter by about $1^{\circ}$ in the horizontal plane, thereby changing the incident angle of the laser pulse from $45^{\circ}$ to $46^{\circ}$. At this angle, the measured transmission and reflection coefficients are $R \simeq 49\%$ and $T \simeq 51\%$, corresponding to an extinction factor $\mathcal{F} = (\delta a)^2 \simeq 4 \times 10^{-4}$ and an amplification factor $\mathcal{A} = 25$, as measured in~\cite{PhysRevA.109.053510}.

From Eqn.s~(\ref{eqn:prompt-out-signal}) and (\ref{eqn:delay-out-signal}), to first order, the extinction factor $\mathcal{F}$ affects both the prompt and delayed interference signals identically. As a consequence, the prompt-delay correlation mechanism used in the HFPNS procedure remains fundamentally independent of the exact operating point of the interferometer and can therefore be validated in the present prototype configuration. The correlated phase-noise subtraction principle demonstrated here is therefore assumed, to first order, to remain valid in the final low extinction DeLLight configuration, although additional technical limitations may arise in the final operating regime, as discussed in Section~\ref{sec:conclusion}.

\subsection{Signal analysis method\label{signal-analysis}}


For each successive CCD recorded image, a barycenter is calculated within a fixed square analysis window, defined as the Region of Interest (RoI) with width $w_{\mathrm{RoI}}$ (white dotted squares in Figure~\ref{fig:sample-image-hfpns}). This procedure is applied to both the prompt and the delayed interference intensity profiles, yielding barycenters $\bar{y}_{\mathrm{P,Sig}}^{\mathrm{ON}}(i)$ and $\bar{y}_{\mathrm{P,Sig}}^{\mathrm{OFF}}(i)$ for the prompt pulse, and $\bar{y}_{\mathrm{D,Sig}}^{\mathrm{ON}}(i)$ and $\bar{y}_{\mathrm{D,Sig}}^{\mathrm{OFF}}(i)$ for the delayed pulse, along the vertical axis, for each successive ON and OFF measurements.

To suppress the noise contributions on the calculated barycenter, we first determine the linear correlation between the prompt and delayed barycenters using only the OFF measurements. The resulting fit parameters, $a_{\mathrm{OFF}}$ and $b_{\mathrm{OFF}}$, characterize the common noise shared by the two pulses. The corrected prompt signal positions are then obtained for each ON and OFF measurement, by subtracting this common noise through a linear regression-based procedure, given by:
\begin{eqnarray} 
\bar{y}_{\mathrm{PD,Sig}}^{\mathrm{OFF}}(i) & = & \bar{y}_{\mathrm{P,Sig}}^{\mathrm{OFF}}(i)   
 - \left( a_{\mathrm{OFF}} \times \bar{y}_{\mathrm{D,Sig}}^{\mathrm{OFF}}(i) + b_{\mathrm{OFF}} \right)  \nonumber \\
\bar{y}_{\mathrm{PD,Sig}}^{\mathrm{ON}}(i) & = & \bar{y}_{\mathrm{P,Sig}}^{\mathrm{ON}}(i)
 - \left( a_{\mathrm{OFF}} \times \bar{y}_{\mathrm{D,Sig}}^{\mathrm{ON}}(i) + b_{\mathrm{OFF}} \right)  
\label{eq:PDcorrections} 
\end{eqnarray}

Finally, the HFPNS-corrected barycenter shift $\Delta y_{\mathrm{HFPNS}} (i)$ for the $i$‑th ON–OFF measurement is obtained by taking the difference between the corrected prompt barycenters of the ON and OFF intensity profiles:
\begin{eqnarray}
\Delta y_{\mathrm{HFPNS}} (i)  =  \bar{y}_{\mathrm{PD,Sig}}^{\mathrm{ON}}(i) -  \bar{y}_{\mathrm{PD,Sig}}^{\mathrm{OFF}}(i).
\label{eq:DeltayHFPNS}
\end{eqnarray}
This approach ensures that common noise between the prompt and delayed pulses is efficiently removed, allowing the extraction of any deflection signal with minimal contamination from the correlated phase fluctuations inside the Sagnac interferometer.

\subsection{Expected shot-noise-limited spatial resolution}

The spatial resolution limited by shot noise has been experimentally characterized in Ref.~\cite{mailliet2024performance}, using a dedicated test bench with the same CCD camera used in the present HFPNS experiment. In that study, the shot-noise-limited spatial resolution, $\sigma_y^{\mathrm{SN}}$, was measured as a function of the peak intensity of the recorded interference beam profile and it is found to be in excellent agreement with corresponding Monte Carlo simulations.

Under the typical operating conditions of the HFPNS setup, the interference signal recorded on the CCD reaches a peak intensity of approximately 1500~ADU. According to the calibration established in Ref.~\cite{mailliet2024performance}, this intensity corresponds to a shot-noise-limited spatial resolution of $\sigma_y^{\mathrm{SN}} \simeq 36$~nm. In the following, this value is taken as the reference shot-noise level.

\section{Results\label{results}}

\subsection{HFPNS Procedure}

For the measurements presented in this article, the pump beam is absent. Consequently, the average value $\langle{\Delta y(i)}\rangle$ of the signal is expected to be zero since there is no interaction between the pump and probe pulses, and the standard deviation of the barycenter signal distribution $\sigma_{\Delta y(i)}$ corresponds to the spatial resolution achieved with the HFPNS setup.

The purpose of these measurements is twofold: first, to validate the performance of the HFPNS method, and second, to identify and characterize the remaining sources of residual phase noise that are not corrected by the system — in particular, the mechanical vibrations induced by the delay stage separating the prompt and delayed pulses. 

\begin{figure}[h]
  \centering
  \includegraphics[width=1\linewidth]{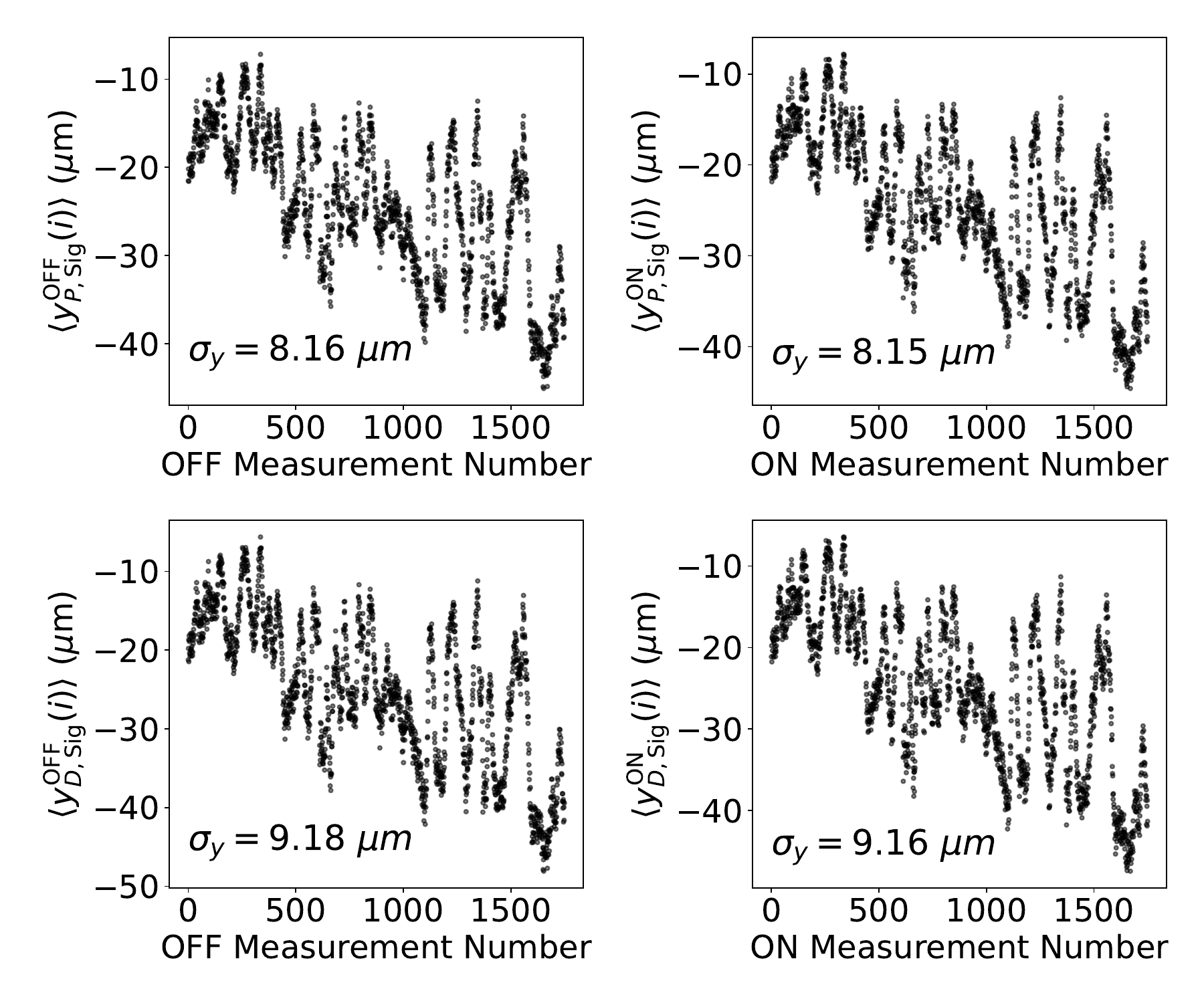}
  \caption{The measured barycenters of the interference intensity profiles within a square analysis window of size $w_{\mathrm{RoI}} = w_{\mathrm{FWHM}}$ for the prompt (upper plots) and delayed (bottom plots) beams, and for the OFF events (left plots) and the ON events (right plots). In all cases, a slow correlated temporal drift is clearly observed across all the measurement sequence.}
  \label{fig:raw-barycenter-plots}
\end{figure}

We present here the measurement of the spatial resolution, obtained with $3500$ successive laser shots (at a $10$~Hz repetition rate). Data of successive odd and even laser shots are arbitrarily separated into OFF and ON data in order to define $i$-th "ON-OFF" measurement at $5$~Hz repetition rate.
The raw vertical barycenters of the interference intensity profiles of the prompt ($\bar{y}_{\mathrm{P}}^{\mathrm{ON}}(i)$ and $\bar{y}_{\mathrm{P}}^{\mathrm{OFF}}(i)$) and delay ($\bar{y}_{\mathrm{D}}^{\mathrm{ON}}(i)$ and $\bar{y}_{\mathrm{D}}^{\mathrm{OFF}}(i)$) are computed within a RoI size equal to the full width at half maximum ($w_{FWHM}$) of the transverse intensity profile. The distributions of the calculated raw barycenters are presented in Figure~\ref{fig:raw-barycenter-plots}, showing large fluctuations of the order of few micrometers. 

\begin{figure}[h]
  \centering
  \includegraphics[width=1\linewidth]{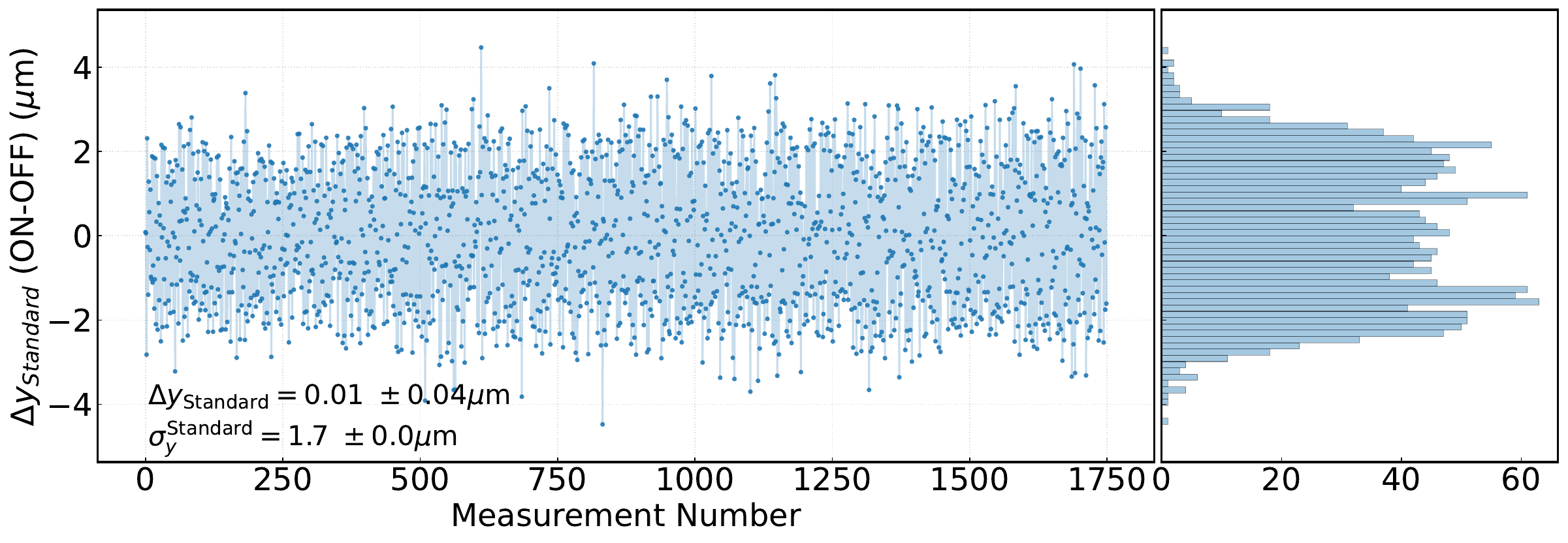}
  \caption{Distribution of the standard ON-OFF signal $\Delta y_{\mathrm{Standard}}$, obtained with direct subtraction between successive ON and OFF barycenter measurements of the interference intensity profile, applied on a sample dataset with 3500 successive laser shots at 10 Hz, corresponding to 1750 ON-OFF measurements. The right panel shows the fitted Gaussian distribution on the obtained data.}
  \label{fig:ON-OFF-result}
\end{figure}

Applying direct subtraction between successive ON and OFF measurements, yields the standard ON-OFF signal
\begin{eqnarray} 
\Delta y_{\mathrm{Standard}}(i)=\bar{y}_{\mathrm{P,Sig}}^{\mathrm{ON}}(i)-\bar{y}_{\mathrm{P,Sig}}^{\mathrm{OFF}}(i)
\end{eqnarray}
The resulting distribution is presented in Figure~\ref{fig:ON-OFF-result} showing that the slow temporal drift is removed. However, large fluctuations ($\sigma_y^{\mathrm{Standard}} = 1.7$~µm) still remain, dominated partly by the beam-pointing instabilities and more importantly by interferometric phase noise.

\begin{figure}[h]
  \centering
  \includegraphics[width=1\linewidth]{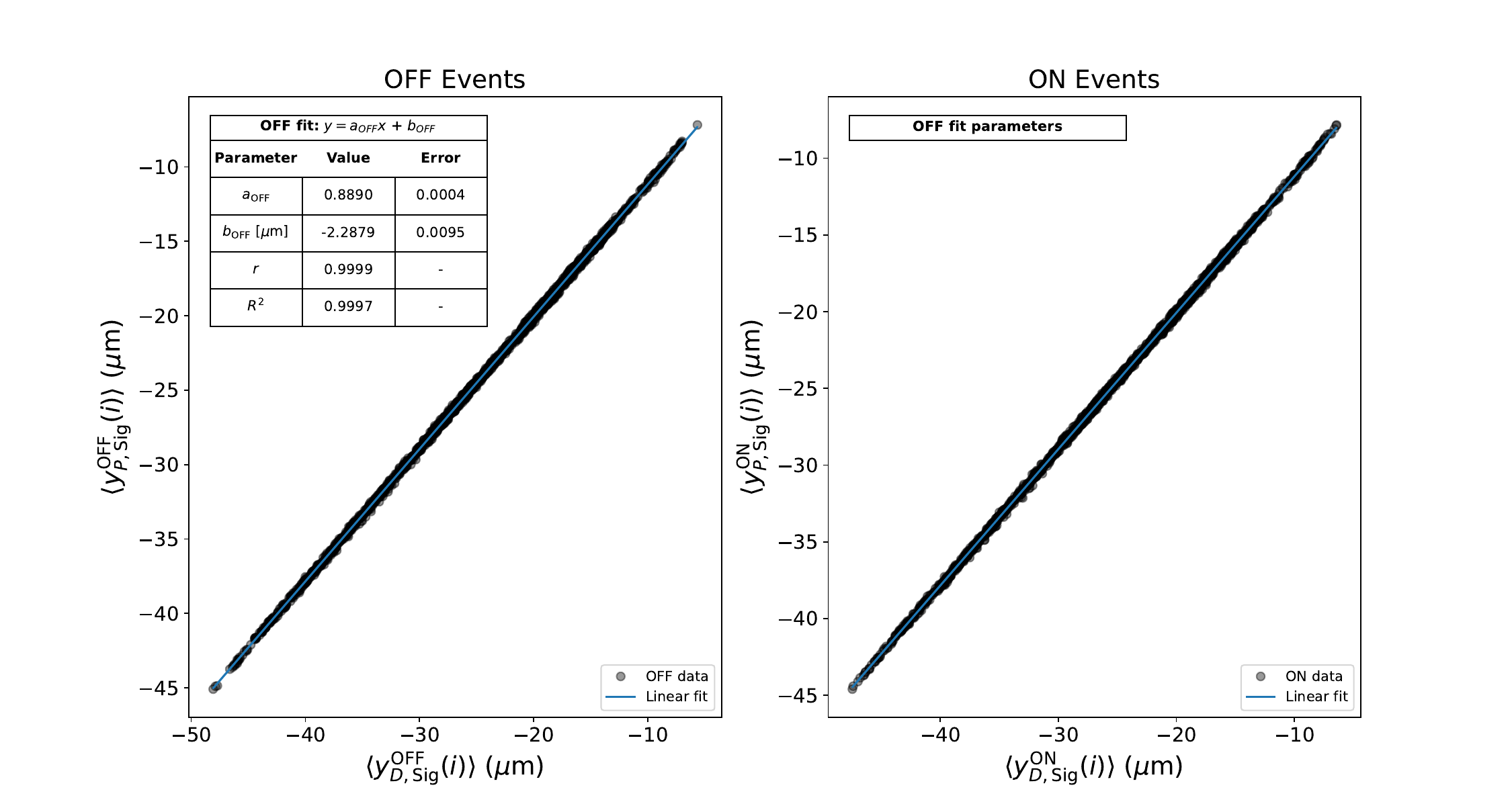}
  \caption{Linear correlation between the prompt and delayed barycenter fluctuations for the OFF and ON datasets. The regression parameters are extracted from the OFF data only and subsequently applied to the complete HFPNS correction procedure. The fitted parameters, uncertainties, Pearson correlation coefficient r, and coefficient of determination $R^2$ are reported in the inset table. The extremely high correlation confirms the common-mode nature of the dominant phase-noise fluctuations between prompt and delay beams.}
  \label{fig:prompt-delay-barycenter-correlation}
\end{figure}

To suppress the noise contributions in the raw barycenter signals of the prompt pulse, the barycenter correlation between prompt and delay beams is used. Figure~\ref{fig:prompt-delay-barycenter-correlation} shows this correlation for the ON and OFF datasets separately. 
A clear linear dependence emerges between the prompt and delayed signals confirming both signals are similarly affected by common noise sources: the beam-pointing fluctuations and the interferometric phase noise.
The linear-regression parameters $a_{\mathrm{OFF}}$ and $b_{\mathrm{OFF}}$ of Eqn.~(\ref{eq:PDcorrections}) are obtained by fitting exclusively the OFF data, and are then used to remove the noise from the prompt signal, using Eqn.~(\ref{eq:PDcorrections}). The HFPNS-corrected signal $\Delta y^{\mathrm{HFPNS}}(i)$ is then calculated, following Eqn.~(\ref{eq:DeltayHFPNS}).

The obtained correlations exhibit extremely high Pearson coefficients $r \approx 0.9999$ together with coefficients of determination $R^2 \approx 0.9997$, confirming the strong common-mode nature of the prompt and delayed fluctuations. The fitted regression parameters obtained from the OFF dataset are $a_{\mathrm{OFF}} = 0.889 \pm 0.00036$ and $b_{\mathrm{OFF}} = -2.288 \pm 0.010 \mu\mathrm{m}$. 

Note that the fitted parameter $a_{\mathrm{OFF}}$ is not exactly equal to 1. 
It reflects the residual asymmetry between the prompt and delayed fluctuations, originating from differences in the corresponding optical paths and residual mechanical contributions of the delay-line section.
To further investigate this deviation of the fitted parameter ($a_{\mathrm{OFF}}$) from unity, the raw barycenter fluctuations of the prompt and delayed pulses were compared before application of the HFPNS correction. As shown in Figure~\ref{fig:raw-barycenter-plots} the prompt pulses exhibit fluctuations of approximately $\sigma_y \approx 8.15~\mu\mathrm{m}$, while the delayed pulses show systematically larger fluctuations of approximately $\sigma_y \approx 9.17~\mu\mathrm{m}$. This relative difference of about $10\%$ is consistent with the observed deviation of the regression slope ($a_{\mathrm{OFF}}$) of about $10\%$ from unity. This behavior suggests that the prompt and delayed pulses do not experience perfectly identical residual mechanical fluctuations and will be readdressed in the following residual noise section.


\begin{figure}[h]
  \centering
  \includegraphics[width=1\linewidth]{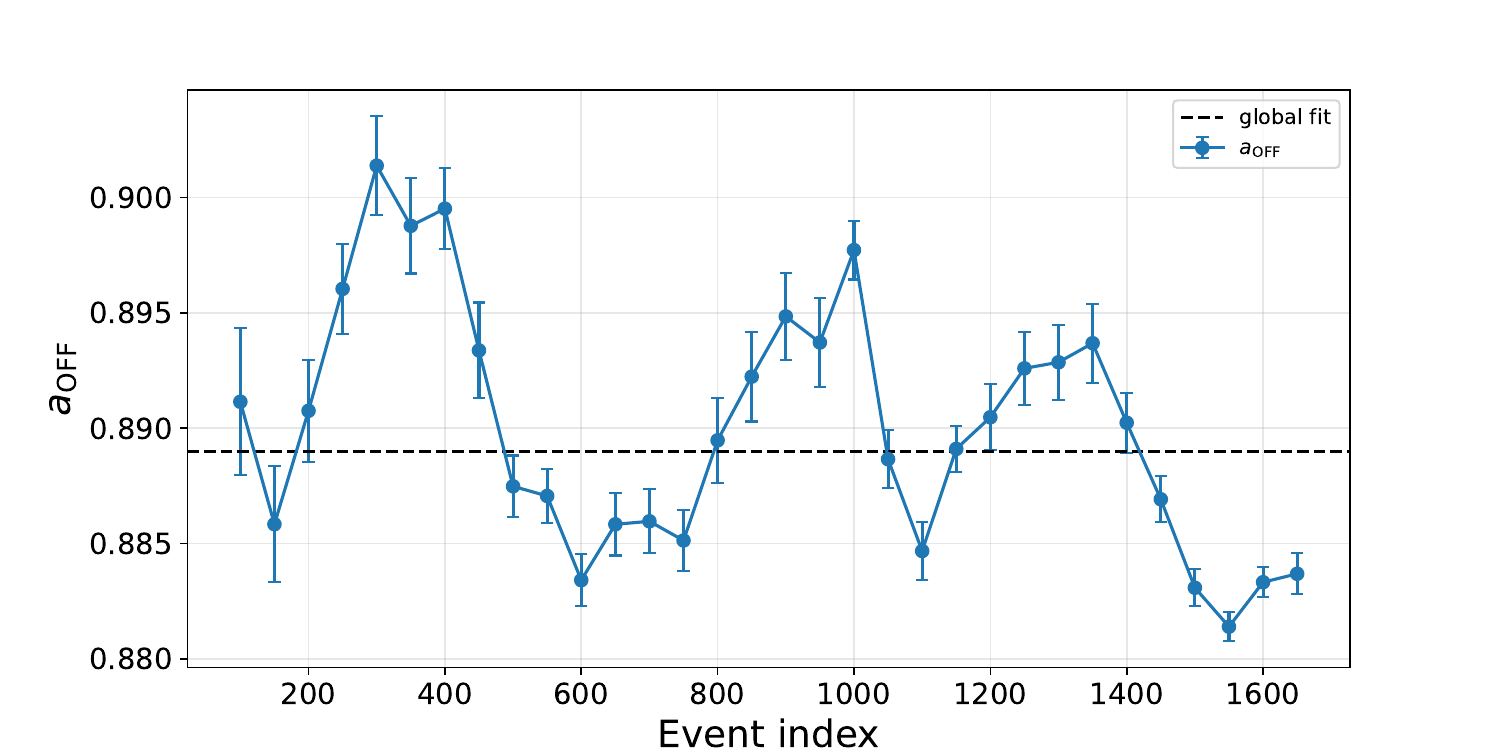}
  \caption{Temporal evolution of the regression parameter $a_{\mathrm{OFF}}$ obtained from sliding-window fits of the full OFF dataset. The dashed horizontal line indicates the global regression value extracted from the complete dataset. The fluctuations of $a_{\mathrm{OFF}}$ remain small throughout the acquisition time, demonstrating the stability of the prompt delay correlation and the robustness of the HFPNS regression procedure.}
  \label{fig:a_off-evolution}
\end{figure}

Further, to investigate the temporal stability of the regression procedure, the fitted parameter ($a_{\mathrm{OFF}}$) was additionally evaluated using a sliding-window analysis over the full dataset. Figure~\ref{fig:a_off-evolution} shows the evolution of the fitted slope parameter throughout the acquisition sequence together with the global regression value obtained from the complete dataset analysis at once. The parameter remains stable over the full acquisition time with fluctuations limited to the percent level around the global mean value. This behavior confirms the robustness of the prompt-delay correlation and indicates that the dominant common-mode fluctuations remain highly correlated throughout the measurement sequence. No significant drift or instability of the regression coefficients is observed over time.

The robustness of the final HFPNS spatial resolution with respect to variations of the regression parameter $a_{\mathrm{OFF}}$ was additionally investigated and is discussed in Appendix~\ref{app:a_off-vs-final-sigma}.

\begin{figure}[h]
  \centering
  \includegraphics[width=1\linewidth]{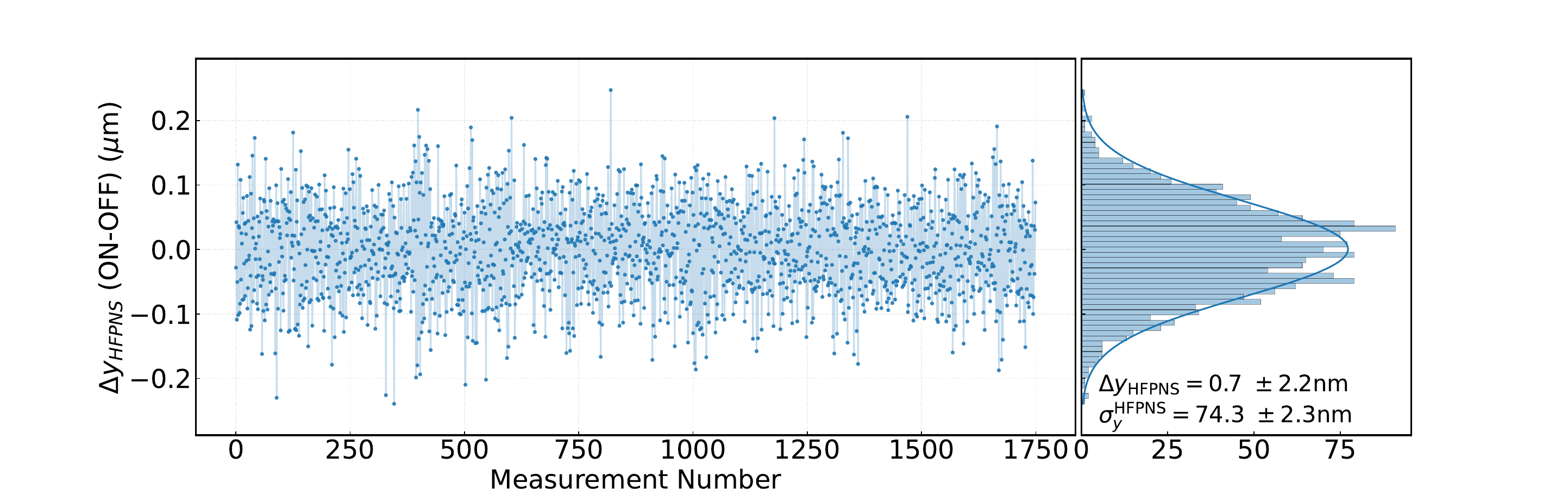}
  \caption{Distribution of the HFPNS ON-OFF signal $\Delta y_{\mathrm{HFPNS}}$, obtained with the HFPNS correction method of the interference intensity profiles. After HFPNS correction, the achieved spatial resolution is  $\sigma_y^{\mathrm{HFPNS}} = 74.3$~nm, and the average signal is $\langle{\Delta y_{\mathrm{HFPNS}}(i)}\rangle =   0.7 \pm 2.2$~nm, compatible with the expected zero value in the absence of pump pulses.}
  \label{fig:standard-vs-hfpns-without-notch}
\end{figure}

The resulting distribution of the HFPNS-corrected ON-OFF signal $\Delta y_{\mathrm{HFPNS}}(i)$ obtained from Eqn.~\ref{eq:DeltayHFPNS} is shown in Figure~\ref{fig:standard-vs-hfpns-without-notch}. The gaussian fit of the $\Delta y_{\mathrm{HFPNS}}(i)$ distribution over all the data collection exhibits a strongly improved spatial resolution, with $\sigma_y^{\mathrm{HFPNS}} = 74$~nm. Compared to the dispersion of the standard ON-OFF signal $\Delta y_{\mathrm{standard}}(i)$ distribution, it corresponds to an improvement by a factor of 28. However, the resulting spatial resolution is still about two times larger than the expected shot-noise-limited resolution.

\subsection{Residual Noise}
The discrepancy between the expected shot-noise limit and the HFPNS-corrected result, referred to here as residual noise, is likely associated with relative difference in the optical path of the prompt and delayed beams, or by instabilities in the delay line both of which can distort the relative intensity profiles prior to their injection into the interferometer. In particular, the delayed pulse propagates through an additional optical delay-line section before entering the Sagnac interferometer and may therefore experience slightly different optical-path fluctuations and mechanical transfer functions compared to the prompt pulse.

This interpretation is further supported by the comparison of the raw prompt and delayed barycenter fluctuations prior to the HFPNS correction. Reminding Figure~\ref{fig:raw-barycenter-plots}, the delayed pulses systematically exhibit larger fluctuations than the prompt pulses, with measured fluctuation amplitudes of $\sigma_y^{\mathrm{Delay}} \approx 9.17~\mu\mathrm{m}$ and $\sigma_y^{\mathrm{Prompt}} \approx 8.15~\mu\mathrm{m},$ respectively. This relative difference of approximately (10$\%$) supports the interpretation that the prompt and delayed pulses experience slightly different residual fluctuations due to differences in their optical paths and mechanical environments.

\begin{figure}[h]
  \centering
  \includegraphics[width=1\linewidth]{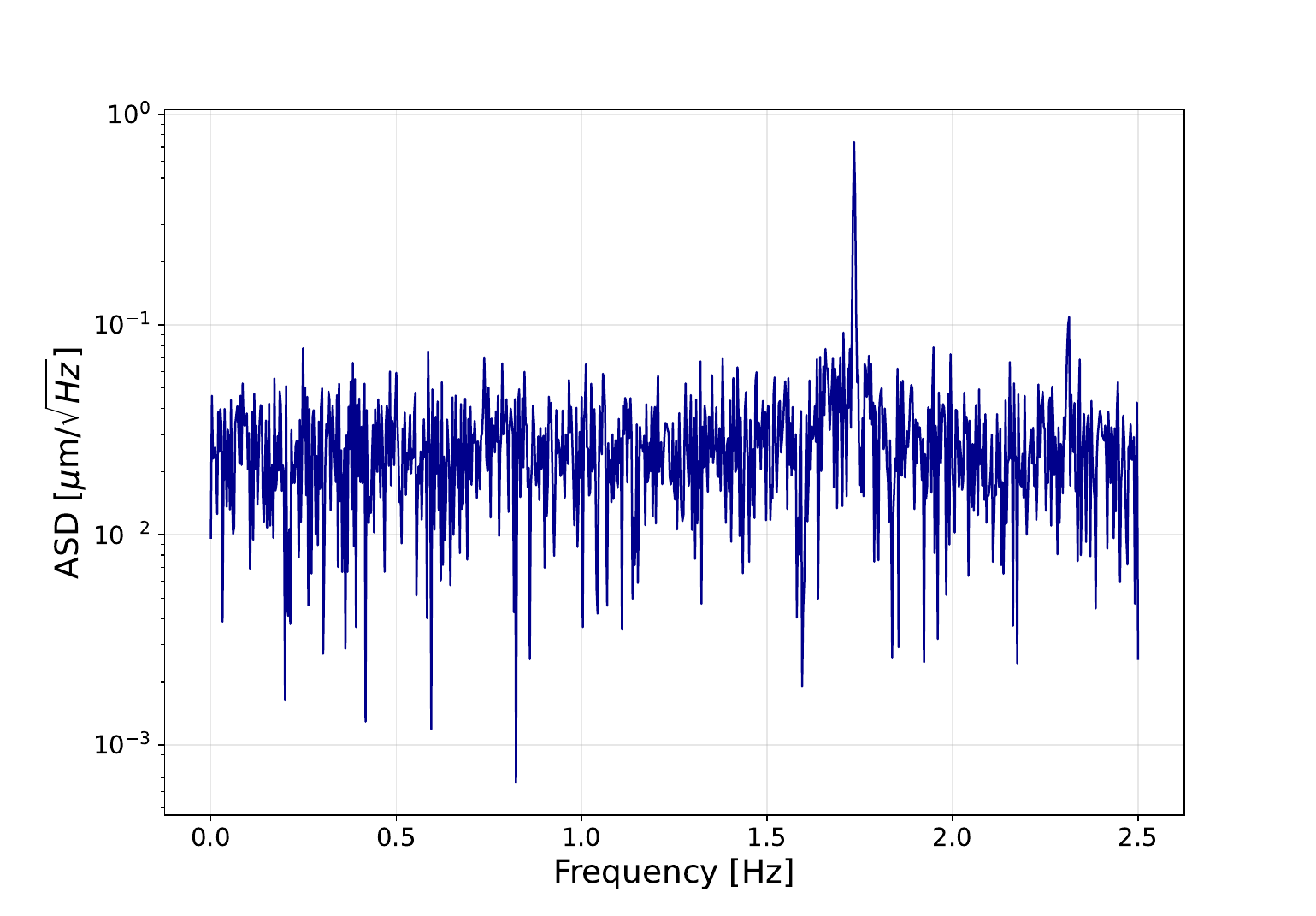}
  \caption{The Amplitude Spectrum Density (ASD) of the HFPNS-corrected interference signal barycenter measurements in logarithmic scale.}
  \label{fig8:PSD-of-HFPNS-Direct}
\end{figure}

To further investigate the origin of these residual fluctuations, the Fourier analysis of the HFPNS-corrected barycenter signal was performed. The resulting amplitude spectral density (ASD), shown in Figure~\ref{fig8:PSD-of-HFPNS-Direct}, exhibits two pronounced narrow-band resonance peaks at distinct frequencies. These spectral features are likely associated with residual mechanical vibrations of the delay-line section or incident beam pointing fluctuations.

\subsection{Study of the nature of the residual noise by using back reflections}

As discussed in~\cite{mailliet2024performance}, back reflections will ultimately be suppressed in the final DeLLight experiment.
However, we propose here to add them in the analysis, in order to investigate the origin of the residual noise and to confirm that it is dominated by the residual relative fluctuations between prompt and delayed beam positions.

As explained in Section \ref{HFPNS}, back reflection spots provide the exact replica of the incident beam. Since they propagate along the same optical path as their corresponding interference signals, the prompt and delayed beams, they carry all beam-pointing fluctuations originating from environmental perturbations and optical elements along the beam line, including those induced by the delay line.

From this perspective, we apply the same HFPNS correction method to the prompt and delay direct back reflection spots, using Eqn.~(\ref{eq:PDcorrections}) applied to the barycenters of the prompt and delay back reflection intensity profile. This procedure yields the prompt–delay (PD) corrected back reflection barycenter measurements, denoted as $\bar{y}_{\mathrm{PD,BR1}}^{\mathrm{OFF}}(i)$, and $\bar{y}_{\mathrm{PD,BR1}}^{\mathrm{ON}}(i)$ for the OFF and ON events, respectively.

\begin{figure}[h]
\centering
\includegraphics[width=1\linewidth]{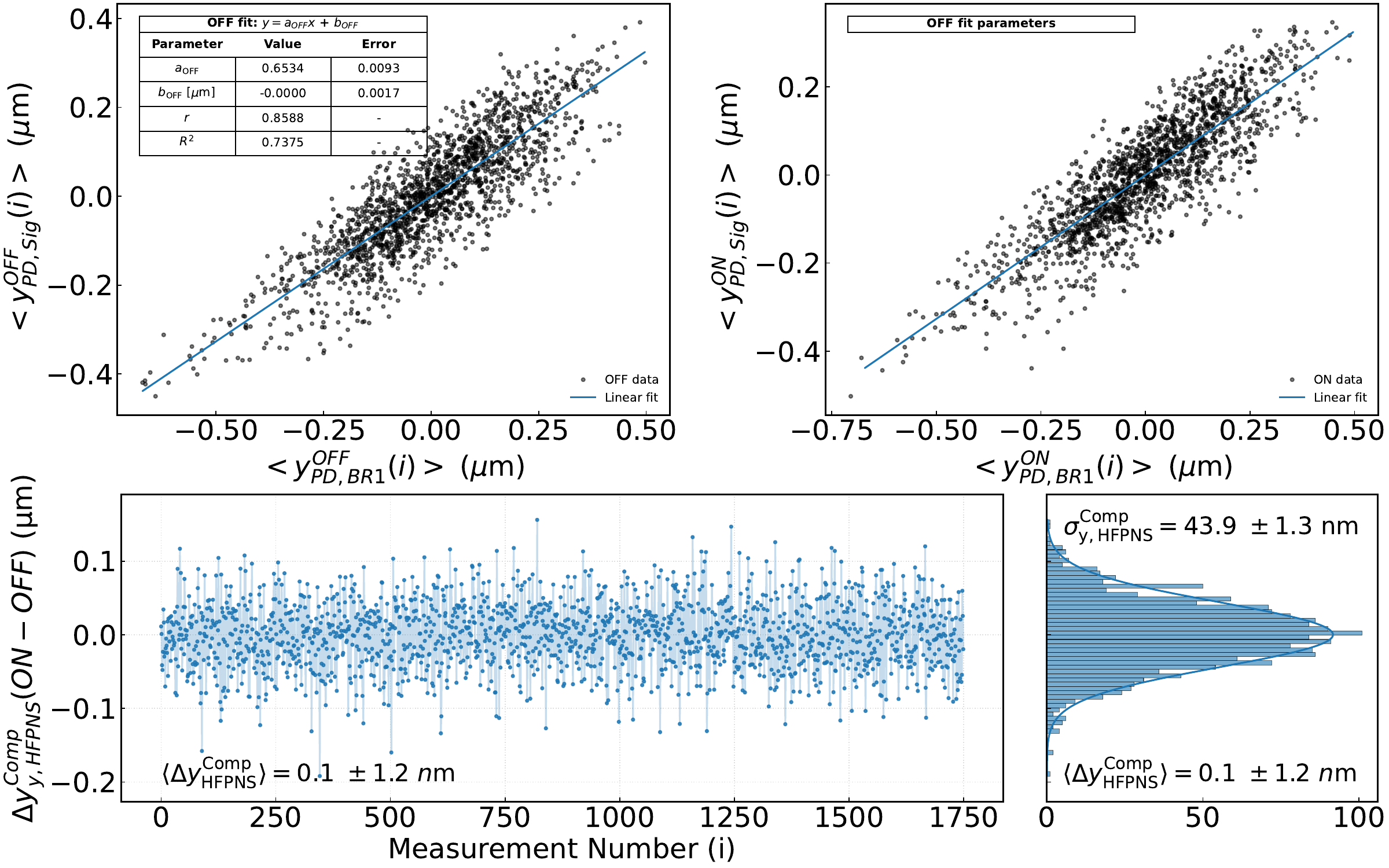}
\caption{Residual correlation between the prompt-delay corrected interference signal and the prompt-delay corrected back reflection signal. (Upper plots) Correlation measured for the OFF and ON datasets. The blue line corresponds to the linear regression obtained from the OFF data. The inset table summarizes the fitted regression parameters together with their statistical uncertainties and correlation metrics. (Lower left) Resulting ON–OFF corrected signal after combining the HFPNS procedure with the additional back reflection correction. (Lower right) Corresponding distribution of the corrected signal. The Gaussian fit gives a spatial resolution of $\sigma_{y,HFPNS}^{Comp}=43.9 \pm 1.3~\mathrm{nm}$ with a mean residual displacement $\langle \Delta y_{HFPNS}^{Comp} \rangle = 0.1 \pm 1.2~\mathrm{nm}$.}
\label{fig9:HFPNS-Complete}
\end{figure}

As shown in Figure~\ref{fig9:HFPNS-Complete}, a clear linear correlation is then observed between the prompt–delay corrected barycenters of the interference intensity profiles and back reflection intensity profiles. This correlation directly results from the relative fluctuations between the prompt and delay beam, induced by the residual delay-line fluctuations. From this perspective, the additional back reflection correction provides useful information concerning the origin of the residual fluctuations remaining after the standard HFPNS correction.

This residual noise is then removed by a linear regression procedure defined with the OFF measurements and applied both for the ON and OFF measurements, given by:
\begin{eqnarray} 
\bar{y}_{\mathrm{PD,Sig,BR}}^{\mathrm{OFF}}(i) & = &  \bar{y}_{\mathrm{PD,Sig}}^{\mathrm{OFF}}(i) - \nonumber \\ 
& & \left( \alpha_{\mathrm{OFF}} \times \bar{y}_{\mathrm{PD,BR}}^{\mathrm{OFF}}(i) + \beta_{\mathrm{OFF}} \right) \nonumber  \\
\bar{y}_{\mathrm{PD,Sig,BR}}^{\mathrm{ON}}(i) & = &  \bar{y}_{\mathrm{PD,Sig}}^{\mathrm{ON}}(i) - \nonumber \\ 
& & \left( \alpha_{\mathrm{OFF}} \times \bar{y}_{\mathrm{PD,BR}}^{\mathrm{ON}}(i) + \beta_{\mathrm{OFF}} \right)  
\label{eq:CompleteHFPNS} 
\end{eqnarray}

Finally, the complete HFPNS-corrected signal for the $i$-th ON–OFF measurement, $\Delta y_{\mathrm{HFPNS}}^{\mathrm{Comp}} (i)$, is obtained after applying ON-OFF subtraction, given by:
\begin{eqnarray}
\Delta y_{\mathrm{HFPNS}}^{\mathrm{Comp}} (i) =  \bar{y}_{\mathrm{PD,Sig,BR}}^{\mathrm{ON}}(i)  - \bar{y}_{\mathrm{PD,Sig,BR}}^{\mathrm{OFF}}(i)
\label{eq:DeltayCompleteHFPNS}
\end{eqnarray}

The distribution of $\Delta y_{\mathrm{HFPNS}}^{\mathrm{Comp}} (i)$ is presented in Figure~\ref{fig9:HFPNS-Complete}. The Gaussian fit gives a spatial resolution of $\sigma_{y, HFPNS}^{Comp}=43.9~\pm~1.3~nm$. This result is in agreement with the expected shot noise $\sigma_{MC} = 44.2~\pm~0.6~nm$ calculated by the Monte-Carlo simulation when including the additional back reflection correction together with relative delay-stage fluctuations of $100$~nrad (rms). This comparison can be primarily used as a diagnostic study of the residual noise origin rather than as the operational correction scheme intended for the final DeLLight configuration.

\begin{figure}[h]
\centering
\includegraphics[width=1\linewidth]{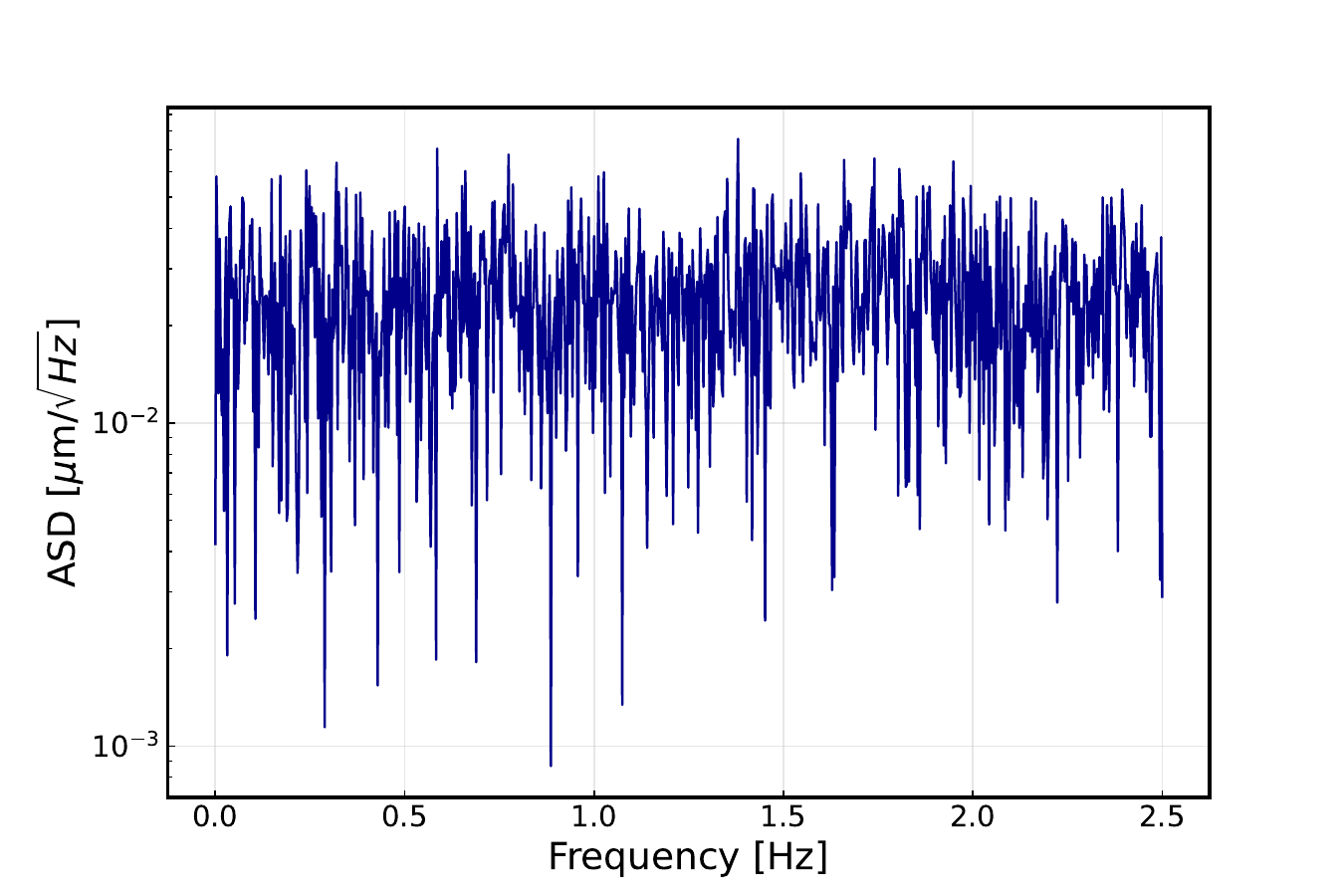}
\caption{ASD of the interference signal obtained after HFPNS correction combined with the additional back reflection correction. No dominant resonance peak is observed after the additional correction procedure.}
\label{fig10:Fourier-HFPNS-Complete}
\end{figure}

The frequency spectrum of $\Delta y_{\mathrm{HFPNS}}^{\mathrm{Comp}} (i)$, shown in Figure~\ref{fig10:Fourier-HFPNS-Complete}, is stable, indicating that the residual noise is predominantly stochastic as expected for the quantum shot noise of the CCD camera.

The shot-noise limit is reached when using the additional back reflection correlations. However, as explained in~\cite{mailliet2024performance}, the final DeLLight configuration requires operation with suppressed back reflections. Indeed, both a small waist of the probe beam at focus and a relatively large focal length are needed to reach a high detection sensitivity. This requires operating with a large beam diameter inside the interferometer and consequently on the beam splitter. As also discussed in~\cite{mailliet2024performance}, a relatively thin beam splitter is additionally required. In such a configuration, the back reflection spots would overlap with the interference signal and would therefore degrade the barycenter measurement.

For these reasons, the back reflections must ultimately be suppressed in the final DeLLight setup by using a beam splitter with enhanced anti-reflective coating. Consequently, only the prompt and delayed interference signals will remain visible at the dark output, motivating the development of an alternative residual-noise suppression strategy compatible with the final operating configuration. This is done in the next section.

\subsection{Suppression of the residual noise by using a digital Notch Filter}

\begin{figure}
  \centering
  \includegraphics[width=1\linewidth]{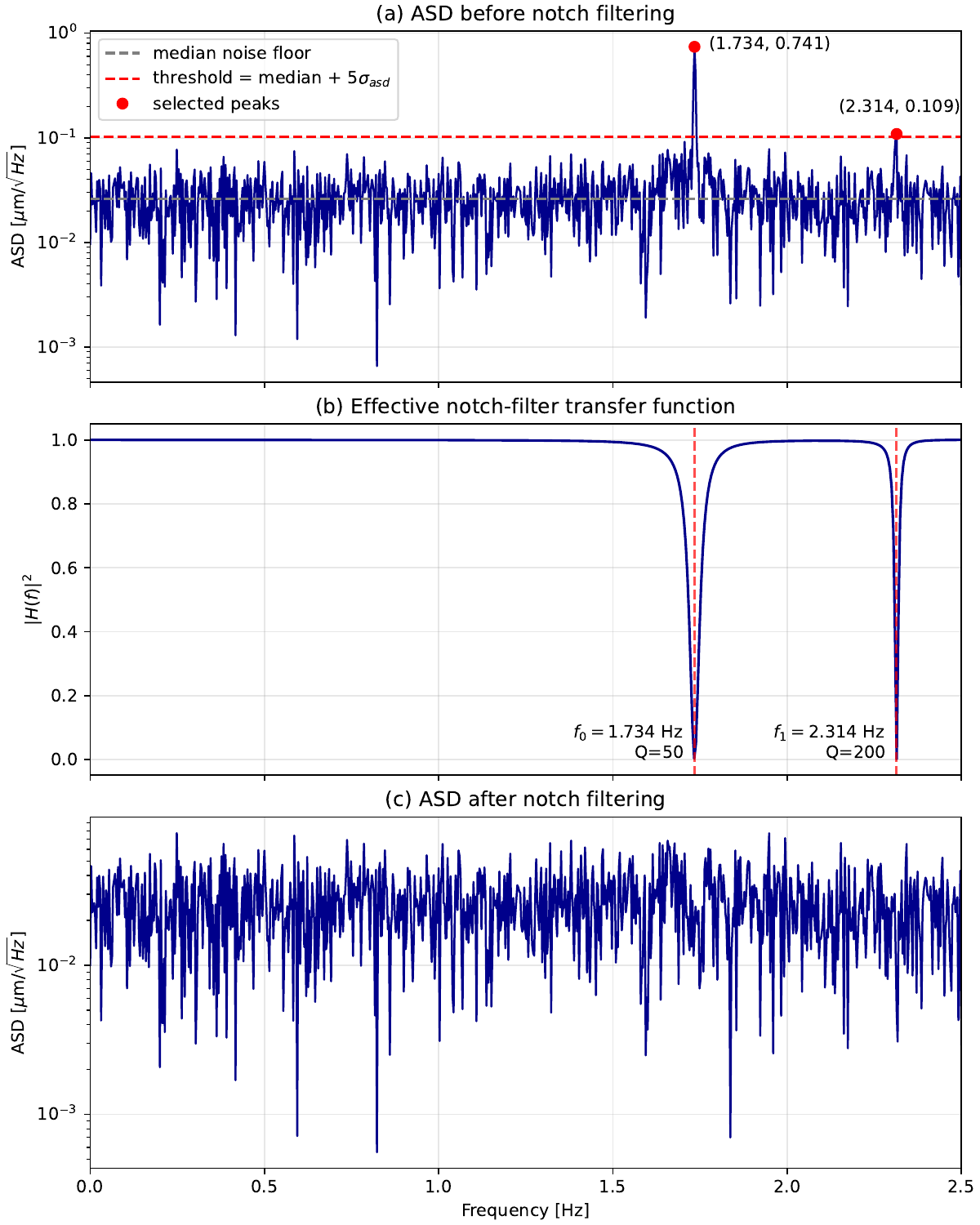}
  \caption{Notch-filtering procedure applied to the HFPNS-corrected signal. (a) Amplitude spectral density (ASD) of the HFPNS-corrected signal before notch filtering. The dashed gray line indicates the median broadband noise floor, while the red dashed line corresponds to the peak-selection threshold defined as the median ASD plus $5\sigma_{asd}$. Two narrow-band resonances are identified at $f_1 = 1.734~Hz$ and $f_2 = 2.314~Hz$ and selected for notch filtering. (b) Effective transfer function $|H(f)|^2$ of the cascaded notch filters applied at the selected resonance frequencies, with quality factors $Q=50$ and $Q=200$, respectively. The filtering is implemented using zero-phase forward-backward filtering (filtfilt) in order to avoid phase distortion of the signal. (c) ASD of the HFPNS-corrected signal after notch filtering. The narrow-band resonances are efficiently suppressed while the broadband noise floor remains essentially unaffected.}
  \label{fig:NotchFiltering}
\end{figure}

We propose here to reject the residual noise by using a numerical notch filter. Recalling the Fourier spectrum of the HFPNS-corrected interference signal shown in Figure~\ref{fig:NotchFiltering}(a), two pronounced resonance peaks are observed. This spectral feature is likely associated with residual mechanical instabilities of the delay line and motivate the use of a dedicated notch filtering approach to suppress this noise contribution in the absence of back reflection signals. 

For this purpose, in the analysis procedure, a digital notch filter is applied on the HFPNS-corrected interference signal. The residual resonances observed after the HFPNS correction are identified through the amplitude spectral density (ASD) analysis of the corrected interference signal. As shown in Figure~\ref{fig:NotchFiltering}(a), the broadband noise floor is first estimated from the median ASD level, and significant spectral peaks are selected using a threshold criterion defined as the median ASD plus $5\sigma_{asd}$. Using this procedure, two dominant resonances are consistently identified at $f_1 =1.734~Hz$ and $f_2 = 2.314~Hz$.

To suppress these narrow-band contributions, two digital notch filters are implemented at the selected frequencies with quality factors $Q=50$ and $Q=200$, respectively. The corresponding effective transfer function $|H(f)|^2$ of the cascaded filters is shown in Figure~\ref{fig:NotchFiltering}(b). The filtering is performed using a forward-backward zero-phase filtering procedure (filtfilt) in order to avoid phase distortion of a potential signal.

The impact of the filtering on the corrected signal is presented in Figure~\ref{fig:NotchFiltering}(c), which shows the ASD after application of the notch filters. The narrow-band resonances are efficiently attenuated while the broadband noise floor remains essentially unchanged. In addition, no structures compatible with obvious aliasing artifacts or filter-induced distortions were observed near the Nyquist frequency.

\begin{figure*}
  \centering
  \includegraphics[width=1\linewidth]{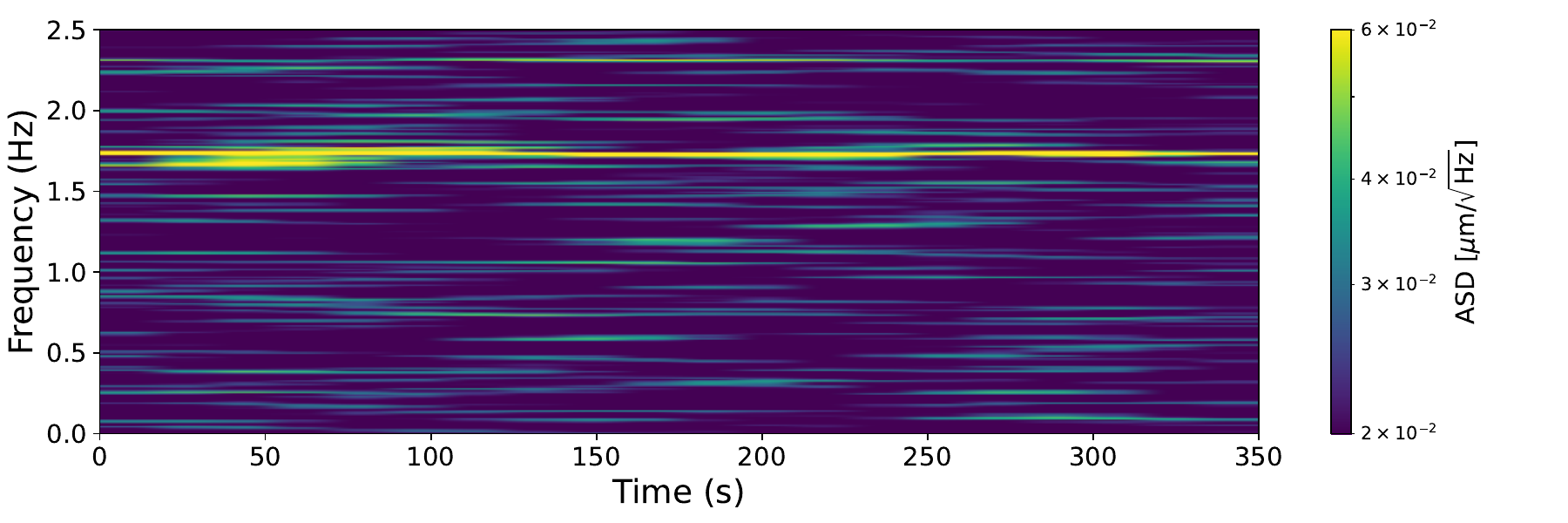}
  \caption{Sliding-window spectrogram of the HFPNS-corrected signal represented as amplitude spectral density (ASD) as a function of time and frequency. The spectrogram is computed using overlapping Hann-windowed Fourier transforms. The dominant resonance at $f_1 = 1.734~Hz$ remains stable throughout the full acquisition time without significant frequency drift or spectral broadening, indicating the stationary narrow-band nature of the noise contribution. A weaker but similarly stable resonance is also observed near $f_2 = 2.314~Hz$. This temporal stability justifies the implementation of fixed-frequency notch filters for the suppression of these narrow-band noise contributions.}
  \label{fig:spectrogram}
\end{figure*}

In addition, to investigate the temporal stability of the identified resonances, a sliding-window spectral analysis was performed on the HFPNS-corrected signal. The dataset was divided into partially overlapping time windows using a Hann window, and the amplitude spectral density was computed independently for each segment. The resulting spectrogram is shown in Figure~\ref{fig:spectrogram}. The dominant resonance at $f_1 = 1.734~Hz$ remains stable throughout the acquisition time without significant frequency drift or broadening, indicating that the observed feature corresponds to a persistent narrow-band mechanical or environmental resonance rather than to transient broadband fluctuations. Similar behavior is also observed for the weaker resonance around $f_2 = 2.314~Hz$, although with lower amplitude. The resonance frequencies remain stable within the experimental frequency resolution throughout the acquisition time. These observations justify the use of fixed-frequency notch filters for the suppression of these narrow-band noise contributions. 

\begin{figure*}
  \centering
  \includegraphics[width=1\linewidth]{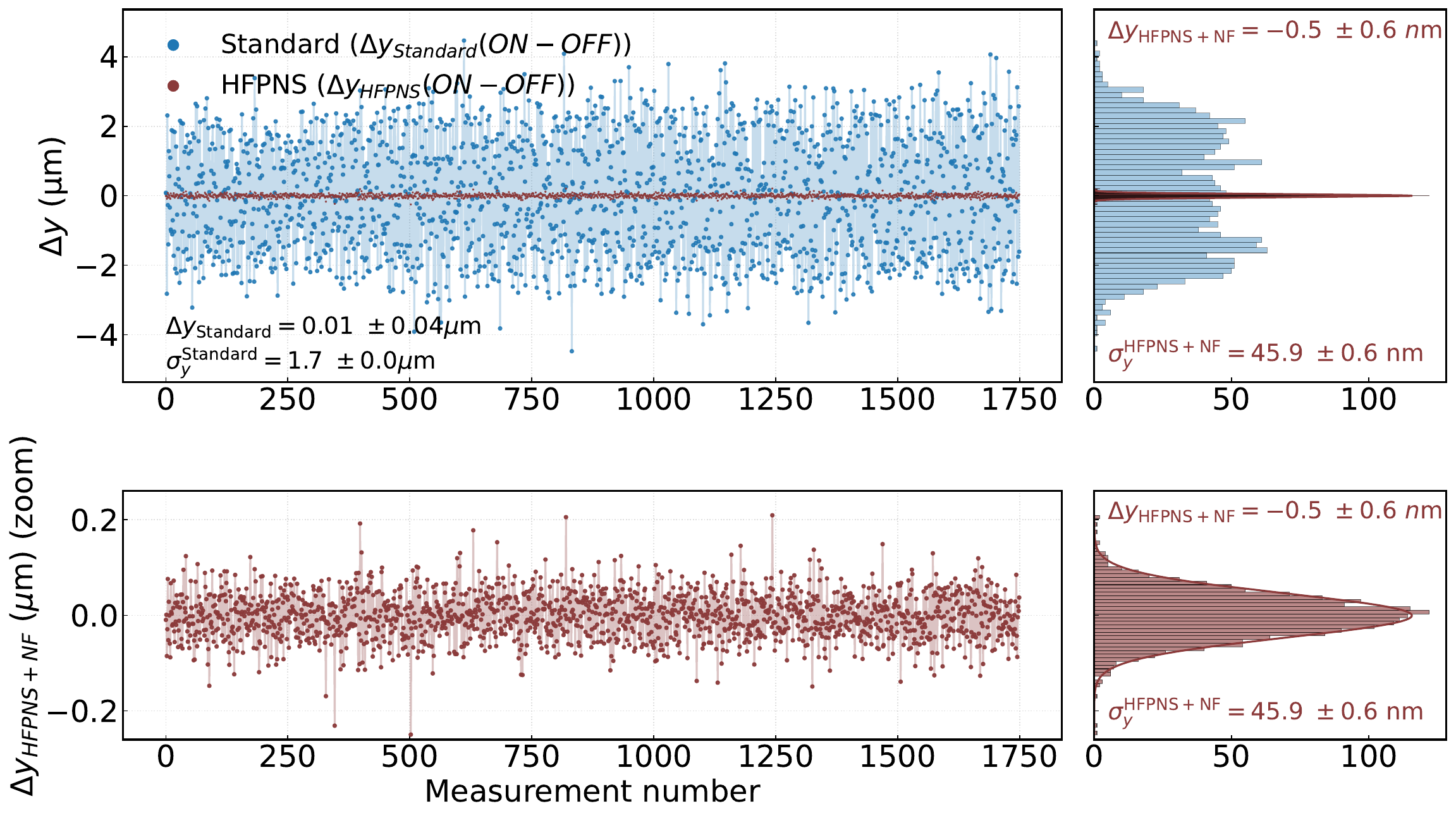}
  \caption{
  Same as Figure~\ref{fig:standard-vs-hfpns-without-notch} but, in red, after applying the HFPNS correction and the numerical Notch filter. 
  After HFPNS and Notch filter corrections, the achieved spatial resolution is  $\sigma_{\mathrm{HFPNS}} = 45.9$~nm, and the average signal is $\langle{\Delta y(i)}\rangle =   0.5 \pm 0.5$~nm, which is compatible with the expected zero value in the absence of pump pulses.
  }
  \label{fig12:final-result}
\end{figure*}

With the notch filter, we obtain the final distribution of the interference barycenter measurement as shown in Figure~\ref{fig12:final-result}, represented by the red data points. A Gaussian fit to this distribution yields a spatial resolution of $\sigma_{y}$ $=$ $45.9 \pm 0.5$ nm. It is only 1.3 times higher than the Monte-Carlo expected quantum-noise-limited spatial resolution of $36$~nm, and indicates that the dominant interferometric phase noise and other sources of noise have been effectively suppressed by the HFPNS method coupled with a numerical notch filter. The average signal is $\langle{\Delta y(i)}\rangle =   -0.5 \pm 0.5$~nm, which is compatible with the expected zero value in the absence of pump pulses.

\begin{figure}
  \centering
  \includegraphics[width=1\linewidth]{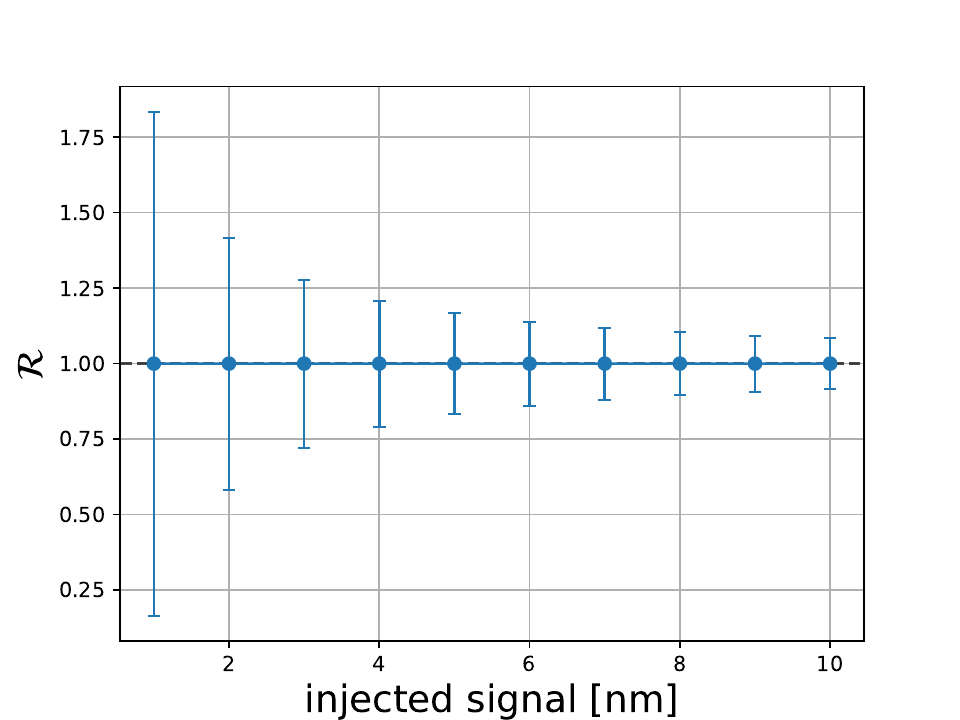}
  \caption{Validation of the signal recovery efficiency using synthetic signal injection. Controlled barycenter shifts with amplitudes ranging from 1 to 10 nm were injected into the ON data only, keeping the OFF data unchanged, and processed through the complete HFPNS and notch-filter analysis chain. The recovery efficiency remains consistent with unity within uncertainties over the explored range, demonstrating that the analysis procedure does not significantly suppress a signal at the expected 5 Hz modulation frequency. The larger uncertainties observed for small injected amplitudes arise from the normalization by the injected signal amplitude and therefore correspond to relative uncertainties.}
  \label{fig:recovered-signal-efficiency}
\end{figure}

To further verify that the analysis chain does not artificially reduce or even suppress a potential physical signal at the expected 5 Hz modulation frequency, an additional synthetic signal injection study was performed. A controlled barycenter shift $\langle \Delta y_{\mathrm{inj}}\rangle$ is injected numerically in the data, by adding it for each ON-OFF $i$‑th measurement to the raw reconstructed barycenter $\bar{y}_{\mathrm{P,Sig}}^{\mathrm{ON}}(i)$ of the prompt signal of the ON event only. In other words, $\bar{y}_{\mathrm{P,Sig}}^{\mathrm{ON}}(i)$ is replaced by $\bar{y}_{\mathrm{P,Sig}}^{\mathrm{ON}}(i) - \langle \Delta y_{\mathrm{inj}}\rangle$ in Eq.~\ref{eq:PDcorrections}, thus reproducing numerically the ON-OFF DeLLight signal.
The complete analysis chain, including the HFPNS linear regression correction and the notch-filtering procedure, is then applied to the modified datasets. The recovered signal is compared to the injected amplitude through the normalized recovery efficiency $\mathcal{R}$, defined as
\begin{equation}
    \mathcal{R} = \frac{\left(\langle \Delta y_{\mathrm{rec}}\rangle - \langle \Delta y_{\mathrm{rec}}^{0}\rangle\right)}{\langle \Delta y_{\mathrm{inj}}\rangle}
\end{equation}
where $\langle \Delta y_{\mathrm{rec}}^{0}\rangle$ corresponds to the average signal of our data without any injected synthetic displacement, which is compatible with the expected zero value due to the absence of pump pulses in the data.
This test is performed with successive amplitudes of the injected signal $\langle \Delta y_{\mathrm{inj}}\rangle$ ranging from 1 to 10~nm. 
As shown in Figure \ref{fig:recovered-signal-efficiency}, the recovery efficiency remains consistent with unity within uncertainties over the explored range, demonstrating that the HFPNS correction and notch filtering do not significantly bias or suppress a signal at the target modulation frequency.

\subsection{Efficiency of HFPNS Method\label{efficiency}}

\begin{figure}
  \centering
  \includegraphics[width=1\linewidth]{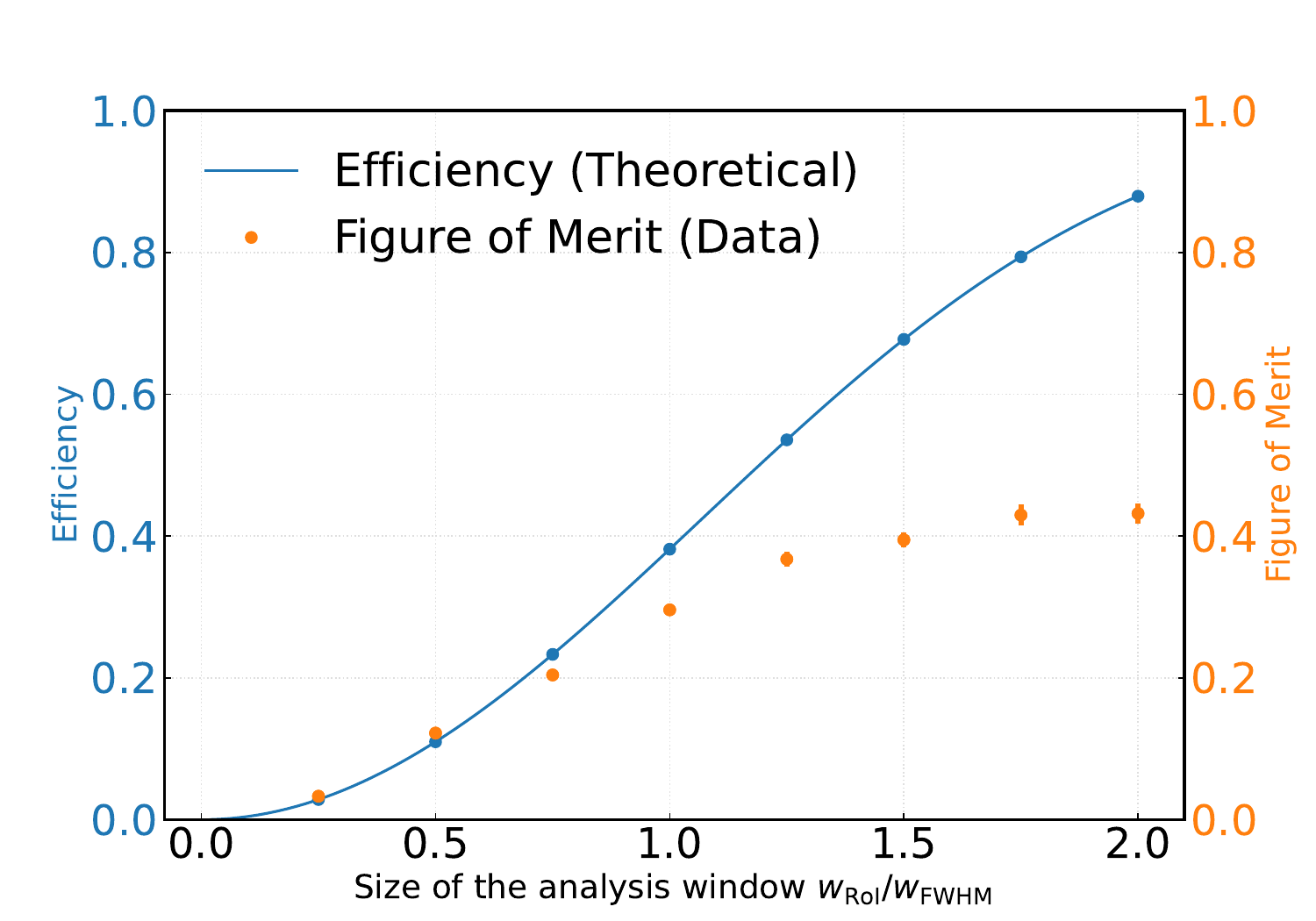}
  \caption{The theoretical efficiency $\epsilon_s$ (blue curve) and the measured figure of merit $\xi$ (orange dots) as a function of the RoI size normalized with the FWHM of the Gaussian beam profile.}
  \label{fig13:eff_and_fom}
\end{figure}

The capacity to measure the displacement of the barycenter of the intensity profile depends on the RoI size of the analysis window. 
We define the efficiency $\epsilon_s(RoI)$ as the ratio between the barycenter shift measured within a given RoI size and the true barycenter shift of the full beam profile. The theoretical efficiency curve, calculated for a Gaussian transverse profile of the beam, is presented in Figure~\ref{fig13:eff_and_fom}, as a function of the RoI size. The efficiency increases with the window size and asymptotically approaches 1 as it becomes large enough to include the entire beam, i.e. $\epsilon_s = 1$ as $w_{RoI} = \infty$.

However, the spatial resolution is strongly degraded when the RoI size is increased because the influence of the interferometric phase noise increases highly with the RoI size , as studied in \cite{mailliet2024performance}. 

\begin{figure}[h]
  \centering
  \includegraphics[width=1\linewidth]{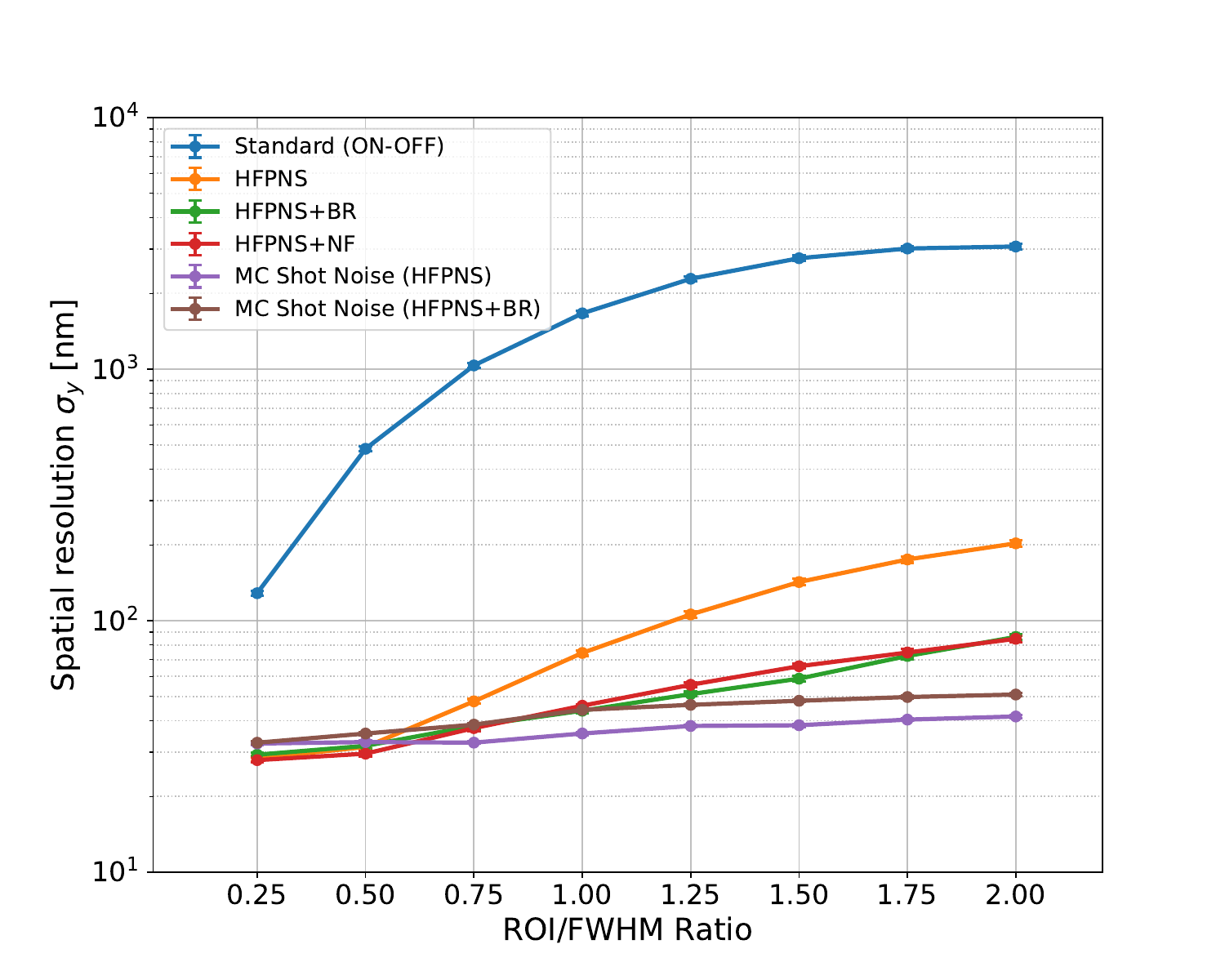}
  \caption{Spatial resolution as a function of the RoI size normalized to the beam FWHM for different analysis procedures. The blue curve corresponds to the standard ON-OFF subtraction of the interference signal barycenter. The orange curve shows the HFPNS correction applied only to the interference signal. The green and red curves correspond to the complete HFPNS correction using additional back reflection correlations and notch filtering, respectively. The violet and brown curves represent the corresponding Monte Carlo-estimated shot-noise limits. Error bars correspond to the statistical uncertainties obtained from the Gaussian fits of corresponding distributions and are smaller than the marker size in several cases.}

  \label{fig14:roi-vs-std}
\end{figure}

Figure~\ref{fig14:roi-vs-std} shows the measured spatial resolution, as a function of the RoI size (given in FWHM unit), for five different analysis methods: standard ON-OFF subtraction (blue), HFPNS method (orange), and HFPNS method with the use of back reflection (green), HFPNS method with notch filter (red), and finally the Monte Carlo shot noise simulation (violet). 

With the standard ON–OFF subtraction, the spatial resolution is strongly degraded as the RoI size is increased, reaching the micrometer scale. This behavior demonstrates the dominance of beam-pointing fluctuations and above all interferometric phase noise, which are not sufficiently suppressed by the ON–OFF procedure and become increasingly dominant in larger window sizes.

In contrast, the HFPNS method result shows significantly improved and more stable spatial resolution over the full range of RoI sizes. The suppression of correlated noise through the prompt–delay correlation leads to a resolution that closely follows the Monte Carlo shot-noise prediction, especially for RoI sizes around the beam FWHM size. 

The results obtained with the additional correction by using back reflection (HFPNS+BR), and applying a passive notch filter to suppress relative peaked noise (HFPNS+NF), indicate that noise contributions which are not directly addressed by the HFPNS method can also be effectively removed. As a result, the measurement sensitivity approaches the ultimate shot-noise limit. This clearly demonstrates that the HFPNS method provides a robust, near shot-noise-limited spatial sensitivity that is weakly dependent on the choice of analysis window size.

Finally, to identify the optimum analysis window (RoI) size for the barycenter measurement, we introduce the figure of merit, $\xi$, defined as the product of the efficiency and the ratio between the shot-noise-limited spatial resolution and the experimentally measured spatial resolution, and given by
\begin{eqnarray}
\xi = \epsilon_s \times \frac{\sigma_{\mathrm{shot~noise}}}{\sigma_{\mathrm{measured}}}
\end{eqnarray}
where $\sigma_{\mathrm{shot~noise}}$ corresponds to the Monte Carlo-estimated shot-noise-limited spatial resolution associated with the considered RoI size. By construction, $\xi = 1$ corresponds to the ideal shot-noise-limited performance, while $\xi < 1$ indicates a degradation of the experimentally achieved sensitivity relative to the ideal shot-noise-limited case.
This parameter accounts for both the signal detection efficiency and the noise performance. For the data presented in this article, the measured figure of merit is presented in Figure~\ref{fig13:eff_and_fom} as a function of the RoI size. It reaches a maximum value of $\xi = 0.44$ for an optimal analysis window size $w_{RoI} = 1.75\times w_{FWHM}$ (with $\epsilon_s=0.8$ and $\sigma_y=75$~nm). 
It corresponds to a residual noise level which is $\xi^{-1}=2.3$ times higher than the ultimate quantum noise limit.

\section{Conclusion}
\label{sec:conclusion}
The results presented in this article confirm the capability of the HFPNS method to efficiently suppress the phase noise arising from fluctuations in the interferometric optics. The method involves introducing a delayed replica of the probe pulse that carries strongly correlated noise characteristics with respect to the prompt pulse, and exploiting the strong linear correlation between the barycenter fluctuations of both pulses. This approach enables efficient noise suppression without compromising the integrity of a potential signal.

The present work corresponds to the experimental validation of the HFPNS methodology on a prototype interferometric configuration without a pump beam. After applying this method on a sample dataset, the spatial resolution obtained, $\sigma_y = 45.9 \pm 0.5$ nm, is close to the CCD camera’s shot-noise limit, around $36~nm$, demonstrating a significant suppression of the correlated interferometric phase noise within the current prototype regime.

Including the efficiency to measure entirely the lateral shift in the intensity profile of the interference signal induced by deflection, which depends on the size of the analysis window, the detectable signal threshold obtained here remains approximately a factor 2.3 above the ultimate shot-noise-limited sensitivity.

The presence of residual uncorrelated noise indicates that certain noise components are not fully measured and subtracted by the current suppression scheme. Additional control measurements based on the back reflection intensity profiles support this interpretation, revealing persistent discrepancies between the prompt and delayed beams that likely arise from optical asymmetries and residual differential fluctuations along the delay path. However, additional technical limitations and residual noise sources may appear in the final high-amplification configuration and are therefore not fully addressed by the present prototype study.

Complementary stabilization techniques or enhanced correction methods, such as optimized notch filtering, may be required to ultimately achieve quantum-limited sensitivity in future implementations. To further enhance the performance of the detection system, several improvements are considered. First, implementing a more advanced notch filtering strategy will allow more efficient isolation and suppression of specific low-frequency noise components. Second, mechanical stabilization of the delay line will be improved to minimize differential fluctuations between the two pulses. Finally, employing a CCD camera with a higher charge storage capacity per unit surface will make it possible to further approach the sensitivity required for future measurements of the QED-predicted optical nonlinearity of the quantum vacuum.

Overall, these results validate the HFPNS approach as a promising methodology for future high-sensitivity interferometric measurements of vacuum optical nonlinearities.

\newpage
\begin{appendices}
\section{Appendix 1}
\label{app:a_off-vs-final-sigma}
\begin{figure}[h]
  \centering
  \includegraphics[width=1\linewidth]{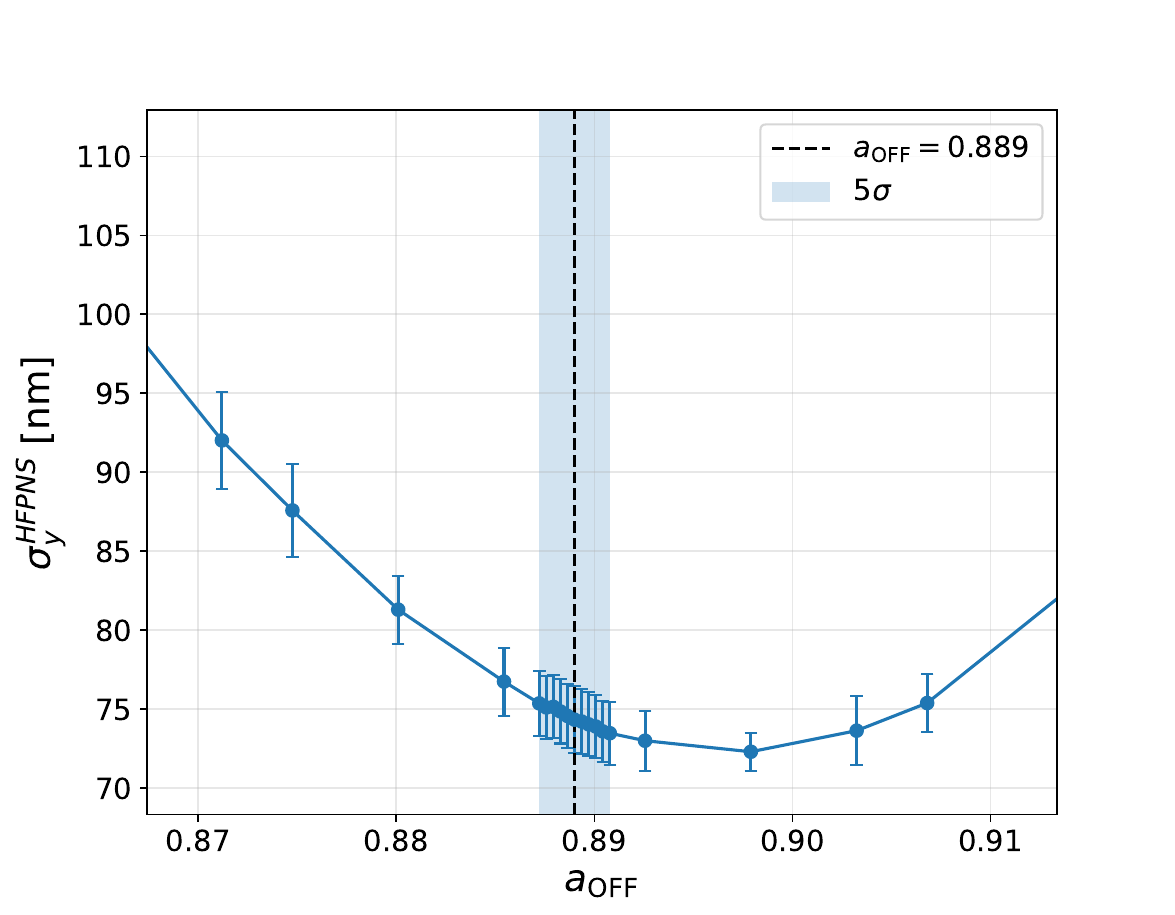}
  \caption{Dependence of the final HFPNS spatial resolution ($\sigma_y^{\mathrm{HFPNS}}$) on the regression parameter ($a_{\mathrm{OFF}}$). The dashed vertical line indicates the nominal regression value extracted from the complete OFF dataset, while the shaded region corresponds to the $\pm5\sigma$ interval associated with the fitted uncertainty on $a_{\mathrm{OFF}}$. The minimum spatial resolution is obtained close to the nominal fitted value, and the final resolution remains only weakly sensitive to variations of the regression parameter within the statistical uncertainty range, demonstrating the robustness of the HFPNS correction procedure with respect to the regression step.}
  \label{fig:a_off_robustness}
\end{figure}

To investigate the influence of the regression uncertainty on the final HFPNS performance, the regression parameter $a_{\mathrm{OFF}}$ in Eqn.~\ref{eq:PDcorrections} was artificially varied around its nominal fitted value over a range corresponding to $\pm 100\sigma_{a_{\mathrm{OFF}}}$, and the complete correction procedure was repeated for each case. Figure~\ref{fig:a_off_robustness} shows the resulting evolution of the final spatial resolution as a function of the imposed regression parameter.

The minimum spatial resolution is obtained close to the nominal value extracted from the OFF dataset. More importantly, the final HFPNS resolution remains only weakly affected by variations of $a_{\mathrm{OFF}}$ within several standard deviations around the fitted value. This behavior indicates that the uncertainty associated with the regression step does not constitute a dominant contribution to the final uncertainty budget and further confirms the robustness of the HFPNS correction procedure.

\end{appendices}

\bibliography{main}

\end{document}